\documentclass[%
 reprint,
superscriptaddress,
longbibliography,
 amsmath,amssymb,
 aps,
prb,
]{revtex4-1}
\setcitestyle{numbers,square}

\usepackage{amsmath}
\usepackage{braket}
\usepackage{graphicx}
\usepackage{dcolumn}
\usepackage{bm}
\usepackage[normalem]{ulem} 
\usepackage{color}
\usepackage{soul}
\usepackage{tabularx}
\usepackage{adjustbox}
\usepackage{hyperref}
\hypersetup{
    colorlinks,%
    citecolor=blue,%
    linkcolor=blue,%
    urlcolor=blue
}

\newcommand{\orcid}[1]{\href{https://orcid.org/#1}{\includegraphics[width=8pt]{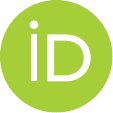}}}

\begin{document}

\title{Group-IV Pentaoctite: A New 2D Material Family}

\author{Vanessa D. Kegler\orcid{0000-0002-6176-2637}}
\email{vanessa.kegler@unir.br}
\affiliation{Physics Graduate Program, Institute of Physics, Federal University of Mato Grosso, Cuiabá, MT, Brazil}
\affiliation{Physics Department, Federal University of Rondônia, Campus Ji-Paraná, Ji-Paraná, RO, Brazil}

\author{Igor S. S. de Oliveira\orcid{0000-0002-9121-9598}}
\email{igor.oliveira@ufla.br}
\affiliation{Physics Department, Federal University of Lavras, Lavras, MG, Brazil}

\author{Dominike Pacine\orcid{0000-0003-0330-7318}}
\email{dominike@iftm.edu.br}
\affiliation{Federal Institute of Education, Science and Technology of Triângulo Mineiro, Uberlândia, MG, Brazil}

\author{Ricardo W. Nunes}
\email{rwnunes@fisica.ufmg.br}
\affiliation{Physics Department, Federal University of Minas Gerais, Belo Horizonte, MG, Brazil}

\author{Teldo A. S. Pereira\orcid{0000-0001-7968-9194}}
\email{teldo@fisica.ufmt.br}
\affiliation{Physics Graduate Program, Institute of Physics, Federal University of Mato Grosso, Cuiabá, MT, Brazil}
\affiliation{National Institute of Science and Technology on Materials Informatics, Campinas, Brazil}

\author{Erika N. Lima\orcid{0000-0002-0670-9737}} 
\email{erika.lima@fisica.ufmt.br}
\affiliation{Physics Graduate Program, Institute of Physics, Federal University of Mato Grosso, Cuiabá, MT, Brazil}
\affiliation{National Institute of Science and Technology on Materials Informatics, Campinas, Brazil}

\date{\today}

\begin{abstract}

This study investigates the structural, mechanical, and electronic properties of novel two-dimensional (2D) pentaoctite (PO) monolayers composed of group-IV elements (PO-C, PO-Si, PO-Ge, and PO-Sn) using first-principles calculations. Stability is explored through phonon spectra and ab initio molecular dynamics simulations, confirming that all proposed structures are dynamically and thermally stable. Mechanical analysis shows that PO-C monolayers exhibit exceptional rigidity, while the others demonstrate greater flexibility, making them suitable for applications in foldable materials. The electronic properties show semimetallic behavior for PO-C and metallic behavior for PO-Si, while PO-Ge and PO-Sn possess narrow band gaps, positioning them as promising candidates for semiconductor applications. Additionally, PO-C exhibits potential as an efficient catalyst for the hydrogen evolution reaction (HER), with strain engineering further enhancing its catalytic performance. These findings suggest a wide range of technological applications, from nanoelectronics and nanomechanics to metal-free catalysis in sustainable energy production.


\end{abstract}

\maketitle

\section{Introduction}

Since the first demonstration of atomically thin two-dimensional (2D) systems, which led to the discovery of graphene through micromechanical cleavage,\cite{novoselov2005,novoselov2004,geim2007} many other 2D materials with structures similar to graphene have been extensively explored in the literature.\cite{liu2019recent} In particular, structures composed of group-IV elements, such as silicene,\cite{cahangirov2009two,silicene-gap} germanene,\cite{cahangirov2009two,germanene-gap} and stanene,\cite{tang2014stable,zhu2015epitaxial} have been theoretically predicted and subsequently synthesized. These materials exhibit properties distinct from graphene, such as topologically nontrivial electronic states.\cite{liu2011quantum,van2014two} Moreover, the presence of a band gap in these materials can overcome graphene's limitations for applications that require a finite band gap, such as the fabrication of field-effect transistors.\cite{schwierz2010graphene}

Numerous 2D planar graphene allotropes composed of carbon atoms have been proposed. Among these, graphyne and graphdiyne, which stand out as two well-studied structures, exhibit unique properties due to acetylenic linkages (consisting of single and triple bonds) between carbon atoms.\cite{kang2018graphyne,li2014graphdiyne} Biphenylene, a planar carbon structure composed of octagonal, hexagonal, and square arrangements of atoms, has recently been synthesized.\cite{fan2021biphenylene} Several other graphene allotropes have been identified, each with distinct features.\cite{enyashin2011graphene,jana2021emerging} These materials possess exceptional electronic and mechanical properties, making them highly relevant for various scientific and technological applications, such as nanoelectronics, energy storage, and biosensors.\cite{bollella2017beyond,peng2014new,zhang2024application} 

Another approach to modifying graphene properties is by introducing extended defects along its structure. For instance, the 558 line defect comprises octagonal and pentagonal $sp^2$-hybridized carbon rings integrated within a pristine graphene lattice.\cite{lahiri2010extended} Moreover, many other defects have been introduced along the graphene structure.\cite{bhatt2022various} These defects can be utilized to tailor the properties of graphene, expanding its range of potential applications.

In Ref.~\onlinecite{PO-C}, the authors propose two new 2D carbon allotropes composed of octagonal and pentagonal carbon rings, demonstrating through first-principles simulations that these structures are energetically and dynamically stable. One structure features a linear arrangement of 558-membered carbon rings, while the other exhibits a zigzag configuration of the same rings. The 558 zigzag configuration has also been shown to form a bismuth 2D sheet, referred to as pentaoctite Bi,\cite{lima2016topologically,erika1} which exhibits topological insulator behavior. Moreover, a new 2D pentaoctite phase in group-V nanostructures, i.e., allotropes of phosphorene, arsenene, and antimonene, has been theoretically investigated.\cite{erika2} The study reveals that these structures are stable, exhibit tunable band gaps, and hold potential applications in optoelectronics due to their absorption in the visible spectrum.

Here, we employ first-principles calculations to explore 2D materials and predict that, in addition to carbon, other group-IV elements (namely silicon, germanium, and tin) can also adopt the pentaoctite structure. We present a detailed investigation of the structural, mechanical, and electronic properties of group-IV pentaoctite sheets. Additionally, we investigate the catalytic activity of the carbon pentaoctite (PO-C) structure in the hydrogen evolution reaction (HER).

\section{Computational details}
 Our first-principles calculations are based on the Density Functional Theory (DFT),\cite{DFT_1, DFT_2} as implemented in the Vienna Ab initio Simulation Package, VASP.\cite{VASP} The generalized gradient approximation (GGA) describes the exchange and correlation potentials as parametrized by Perdew, Burke, and Ernzerhof (PBE).\cite{GGA} The interactions between the valence electrons and the ionic cores are treated using the projector augmented wave (PAW) method.\cite{PAW_1, PAW_2} The electronic wave functions are expanded on a plane-wave basis with an energy cutoff of 500 eV. The supercells are on the $xy$-plane and the Brillouin zone integrations are performed using a 10$\times$10$\times$1 $\Gamma$-centered Monkhorst-Pack sampling.\cite{Monkhorst} 
To avoid interactions between the periodic images of the supercell, the systems are modeled using supercells repeated periodically on the $xy$-plane with a vacuum region of about 20~\AA\ along the $z$-direction. The cell parameters and ionic positions are fully optimized until the residual force on each atom is less than 0.01~eV/\AA. Phonon properties were calculated using the Density Functional Perturbation Theory (DFPT) method, as implemented in the PHONOPY code.\cite{Phonopy} The thermal stability was verified by \textit{Ab Initio} Molecular dynamics (AIMD) simulations using the Andersen thermostat.\cite{andersen} The elastic constants, C$_{ij}$, were calculated based on the stress-strain method implemented in the VASPKIT code.\cite{VASPKIT} In the electronic structure calculations, we used the Hyed-Scuseria-Enzerhof (HSE06) hybrid functional\cite{HSE06} to avoid underestimating the band gap value of the group-IV pentaoctite monolayers. The screened parameter was set to 0.2 \AA$^{-1}$, and 25\%  of the screened Hartree-Fock (HF) exchange was mixed with the PBE exchange functional in the HSE06 method. Additionally, all electronic structure calculations include spin-orbit coupling (SOC). 
 
\section{Results and Discussion}
\subsection{Structural properties and stabilities}
\begin{figure}[!htb]
    \centering
    \includegraphics[width=1.0\linewidth]{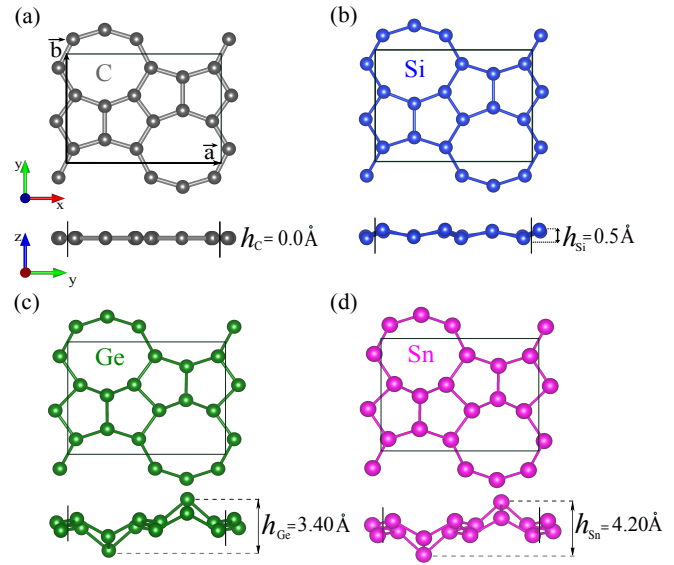}
    \caption{Top and side views of the crystal structures of the (a) PO-C, (b) PO-Si, (c) PO-Ge, and (d) PO-Sn monolayers. The primitive cell containing 12 atoms is marked in the black rectangular area with solid lines defined by the lattice vectors $\Vec{a}$ and $\Vec{b}$. \textit{h}$_{C}$, \textit{h}$_{Si}$, \textit{h}$_{Ge}$ and \textit{h}$_{Sn}$ represent the buckling height for PO-C, PO-Si, PO-Ge, and PO-Sn monolayers. The grey, blue, green, and magenta balls represent the C, Si, Ge, and Sn atoms.}
    \label{fig:estruturas}
\end{figure}

The pentaoctite (PO) monolayers composed by elements from the group-IV
(PO-IV, for IV = C, Si, Ge, and Sn) crystallize in the orthorhombic symmetry with 12 atoms in the unit cell, as shown in the structural models in Fig.\ref{fig:estruturas}.
The top views reveal that these new 2D allotropes are composed of two side-sharing pentagons connected to an octagon. The bond lengths vary from 1.39–1.48 \AA~ for PO-C, 2.24–2.31 \AA~ for PO-Si, 2.44–2.53 \AA~ for PO-Ge, and 2.83–2.92 \AA~ for PO-Sn monolayers. The side view in Fig.\ref{fig:estruturas}(a) reveals that the PO-C sheet
presents a planar structure similar to graphene. On the other hand, the remaining structures present a buckled construction [see Figs.\ref{fig:estruturas}(b),\ref{fig:estruturas}(c) and \ref{fig:estruturas}(d)], which is similar to their group-IV graphene-like counterparts,\cite{cahangirov2009two,tang2014stable} and also to group-V pentaoctite.\cite{erika2}
It is worth mentioning that PO-Si shows a simple two-layered atomic structure, while PO-Ge and PO-Sn have a more complex three-layered structure. It occurs because the buckling height (\textit{h}) follows the increases in the atomic radius of group-IV.
The optimized structural parameters and bond lengths of the PO-IV structures are presented in Table \ref{tab:parameters}.


\begin{ruledtabular}
\begin{table}[ht]
\caption{\label{tab:parameters} Optimized structural parameters of PO-IV monolayers, including lattice constants (\textit{a} and \textit{b}), buckling height (\textit{h}), and bond length ranges.}

\begin{tabular}[t]{lcccc}
Structure & \textit{a} (\AA)& \textit{b} (\AA)  & \textit{h} (\AA) & Bond length range (\AA)\\
\hline
PO-C &6.9&4.87&0.00&1.39--1.48\\
PO-Si &10.81&7.68&0.51&2.24--2.31\\
PO-Ge &10.56&7.86&3.40&2.44--2.53\\
PO-Sn &11.85&8.98&4.20 &2.83--2.92 \\
\end{tabular}
\end{table}
\end{ruledtabular}

\begin{figure}[!htb]
    \centering
    \includegraphics[width=1.0\linewidth]{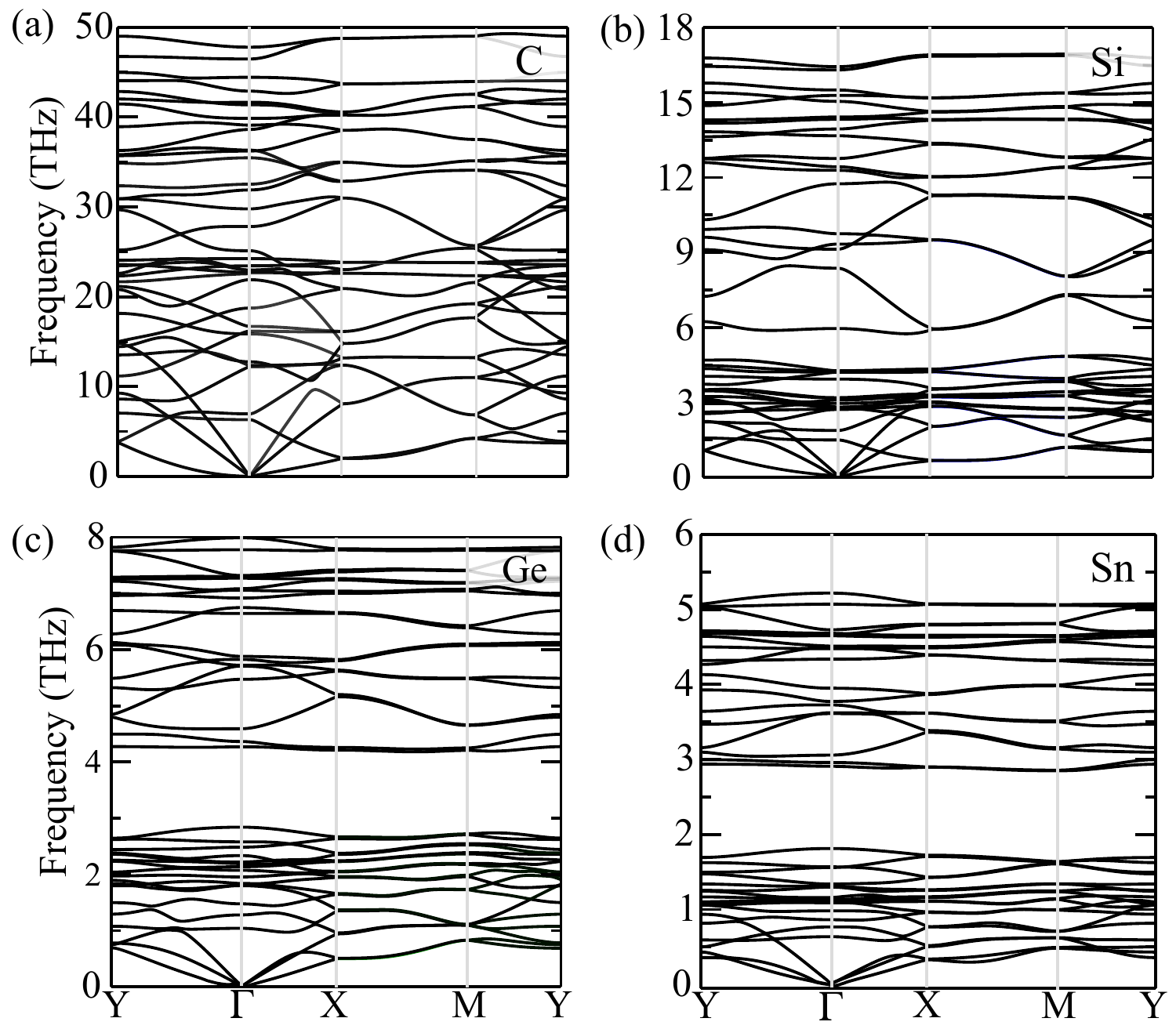}
    \caption{Phonon dispersion relations of the (a) PO-C, (b)PO-Si, (c)PO-Ge, and (d) PO-Sn monolayers.}
    \label{fig:phonons}
\end{figure}
\begin{figure}[!htb]
    \centering
    \includegraphics[width=1.0\linewidth]{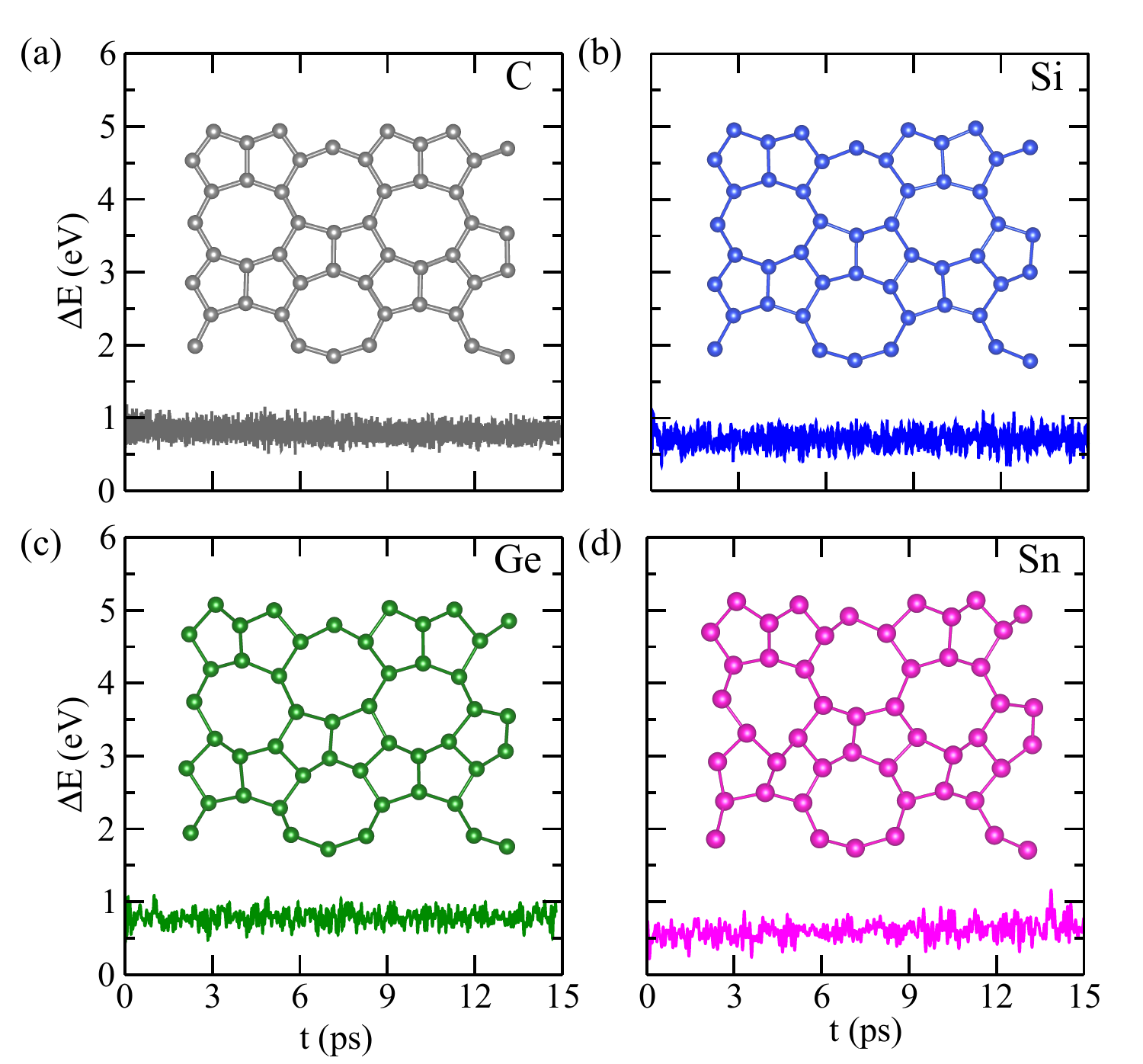}
    \caption{Potential energy fluctuation profiles during AIMD simulations at 300 K for the (a) PO-C, (b) PO-Si, (c) PO-Ge, and (d) PO-Sn monolayers. The snapshots of the crystal structures at 15 ps for the monolayers are shown in the insets.}
    \label{fig:dinamica}
\end{figure}

Fig.\ref{fig:phonons} (a)-(d) shows the phonon dispersion of the PO-IV monolayers. No imaginary frequencies are observed within the first Brillouin zone, indicating its dynamical stability. Moreover, near the $\Gamma$ point of the Brillouin zone, the lowest frequency mode (ZA) exhibits quadratic dispersion. In contrast, the two other modes, longitudinal acoustic (LA) and transverse acoustic branches(TA), display a linear dependence on the wave vector, which is characteristic of 2D materials. Then, we check the structural stability of these PO-IV monolayers at room temperature through \textit{ab initio} molecular dynamics (AIMD) simulations. Fig.\ref{fig:dinamica} (a)-(d) displays the potential energy fluctuations ($\Delta$E) during the simulation time, along with a crystal structure snapshot of the last configuration at 15 ps. All structures maintain their integrity at temperatures up to 300 K, demonstrating good thermal stability at room temperature.


We also verified the energetic stability of these 2D allotropes by calculating their formation energy (E$_F$) relative to their corresponding hexagonal structures. To compute the (E$_F$), we used the equation E$_{F}$=E$_{PO}$-E$_{hex}$, where E$_{PO}$  represents the energy per atom in a PO structure and E$_{hex}$ represents the energy per atom in a hexagonal structure. The calculated relative formation energies for C, Si, Ge, and Sn structures are 350, 120, 90, and 60 meV/atom. These results confirm that the PO-IV structure is energetically metastable compared to its previously synthesized hexagonal counterparts.\cite{novoselov2004,Si-Ge,Sn} Furthermore, when we compared these values with another previously synthesized carbon allotrope, biphenylene (E$_F$ = 470 meV/atom),\cite{biphenele-sint} it is clear that this new 2D allotrope of group-IV has significant potential for experimental realization.

\subsection{Mechanical properties}

From the above indications that the PO-IV is metastable, we proceeded to analyze its mechanical stability.  The elastic constants C$_{11}$, C$_{22}$, $\rm{C}_{12}=\rm{C}_{21}$, and C$_{66}$ and the mechanical properties of this new 2D rectangular lattice allotrope are listed in Table \ref{tab:constants}, together with the values obtained from hexagonal structures. The elastic constants satisfy the Born-Huang criterion:\cite{born1,born2}
$\rm{C}_{11}\rm{C}_{12} - \rm{C}_{12}^2>0$ and $\rm{C}_{66}>0$, 
suggesting that PO-IV sheets are mechanically stable to resist small deformations. 

\begin{table*}
\caption{\label{tab:constants} Elastic constants (C$_{11}$, C$_{12}$, C$_{22}$, C$_{66}$), Young's modulus (E$_x$, E$_y$, E$_{max}$), and Poisson’s ratio ($\nu_x$, $\nu_y$, $\nu_{max}$) of the PO-IV monolayers. The elastic constants and Young’s modulus are given in units of N/m, while Poisson’s ratio is dimensionless.}
\begin{ruledtabular}
\begin{tabular}{cccccccccccc}
Structure &  C$_{11}$  & C$_{22}$ & C$_{12}$ & C$_{66}$ &  E$_x$ & E$_y$ & E$_{max}$ & $\nu$$_x$& $\nu$$_y$  & $\nu_{max} $ & Reference\\
\hline 
PO-C & 270.21 & 345.22 & 65.14 & 100.61 & 333.01 & 254.50 & 333.01 & 0.19 & 0.24 & 0.29 & This work\\
Graphene & 354.12 & 354.12 & 58.65 & 142.38 & 344.41 & 344.41 & 344.01 & 0.17 & 0.17 & 0.17 &\cite{constant-C}\\
PO-Si & 65.95 & 47.52 & 17.96 & 16.54 & 56.61 & 42.40 & 56.61 & 0.38 & 0.28 & 0.38 & This work\\
Silicene & - & - & - & - & 61.33 & 61.33 & 61.33 & 0.32 & 0.32 & 0.32 &\cite{constants-Ge-Sn}\\
PO-Ge & 8.50 & 28.09 & 2.61 & 4.66 & 8.26 & 27.28 & 27.28 & 0.09 & 0.31 & 0.31 & This work\\
Germanene & - & - & - & - & 42.05 & 42.05 & 42.05 & 0.33 & 0.33 & 0.33 &\cite{constants-Ge-Sn}\\
PO-Sn & 5.82 & 14.04 & 1.23 & 0.87 & 5.81 & 13.60 & 13.60 & 0.10 & 0.20 & 0.70 & This work\\
Stanene & - & - & - & - & 24.46 & 24.46 & 24.46 & 0.39 & 0.39 & 0.39 &\cite{constants-Ge-Sn}\\
\end{tabular}
\end{ruledtabular}
\end{table*}

Our results indicate a decrease in Young's modulus (E$_{max}$) of PO-IV structures from C to Sn (333.01 -- 13.60 N/m), following the same tendency observed in hexagonal phases (344.01 -- 24.46 N/m). This trend in Young's modulus directly results from the transition from strong covalent bonding in small atoms (C) to weaker metallic bonding in larger atoms (Sn). The decrease in bond strength as we move down the group-IV elements is a crucial factor in the stiffness of the materials, leading to a lower Young's modulus. The elastic moduli values of the PO-C monolayer are E$_x$ = 333.01 N/m  and E$_y$ = 254.50 N/m. It is important to note that in the $x$-direction, Young's modulus of PO-C is only 3.31\% lower than that of graphene (344.41 N/m).\cite{constant-C} Furthermore, compared to biphenylene, which has Young's moduli of E$_x$ = 259.7 N/m and E$_y$ = 212.4 N/m,\cite{bipheneleno-sci} PO-C exhibits a 27.45\% larger modulus in the $x$-direction and 19.82\% larger along the y-direction. These findings suggest that PO-C is highly robust, making it an ideal material for nanomechanical applications due to its exceptional intrinsic strength.

\begin{figure*}[ht!]
\includegraphics[width=13cm,scale=0.1,clip, keepaspectratio]{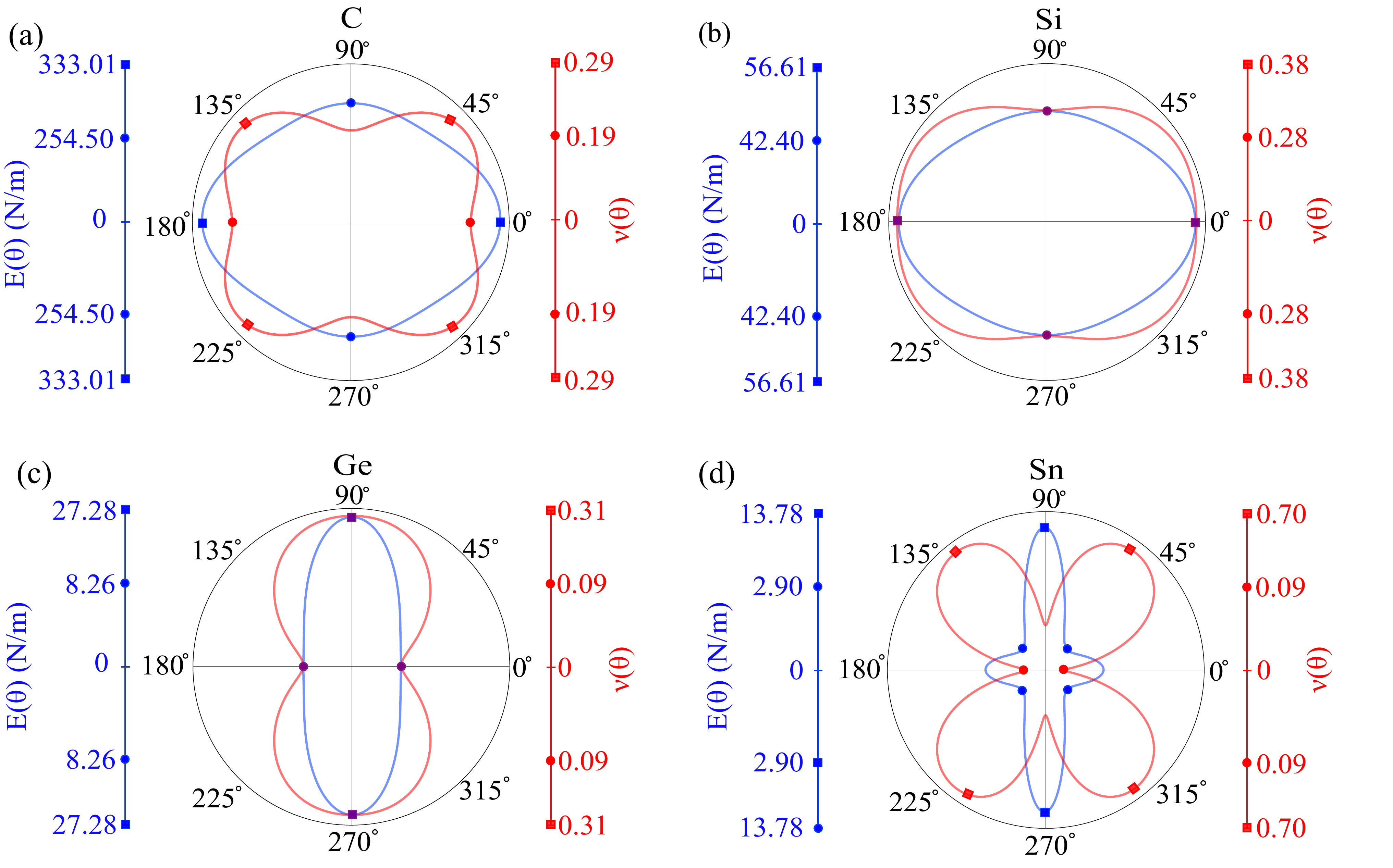}
\caption{\label{fig:moduli} Polar plots illustrating the directional dependence of Young's modulus and Poisson's ratio of the (a) PO-C, (b) PO-Si, (c) PO-Ge, and (d) PO-Sn monolayers. The squares (circles) indicate
the maximum (minimum) values of E($\theta$) (blue symbols) and $\nu$($\theta$) (red symbols).} 
\end{figure*}

The elastic moduli of Si (E$_x=56.61$~N/m, E$_y=42.40$~N/m), Ge (E$_x=8.26$~N/m, E$_y=27.28$~N/m), and Sn (E$_x=5.81$~N/m, E$_y=13.60$~N/m) sheets are also smaller than those of their respective hexagonal phases, as listed in Table \ref{tab:constants}, as well as other well-known 2D materials, such as MoS$_2$ (120 N/m)\cite{MoS2} and black phosphorus (E$_{x}= 22$~N/m, E$_{y}=56$~N/m).\cite{Black-phosphorus} Thus, we can infer that Si, Ge, and Sn structures hold great potential for applications in folding materials.

\begin{eqnarray}
   \label{eq.E}
   E(\theta)&=&\frac{X}{C_1a^{2}+[X/C_{66}-2C_{12}]ab+C_{22}b^{2}}, \\
  \label{eq.nu}
   \nu(\theta)&=&\frac{C_{12}a^{2}-[C_{11}+C_{22}-X/C_{66}]ab+C_{12}b^{2}}{C_1a^{2}+[X/C_{66}-2C_{12}]ab+C_{22}b^{2}},
\end{eqnarray} 
where $a=\sin^{2}(\theta)$, $b=\cos^{2}(\theta)$, and $X=C_{11}C_{22}-C_{12}^{2}$. According to the diagrams in Fig.\ref{fig:moduli}, Young’s modulus and Poisson’s ratio exhibit strong anisotropy across all PO-IV structures, with the effect being more pronounced in the PO-Ge [fig.\ref{fig:moduli}(c)] and PO-Sn [fig.\ref{fig:moduli}(d)] monolayers. For PO-C and PO-Si, the maximum values of Young’s modulus (333.01 and 56.61 N/m) occur at $\theta$ = 0\textdegree, and the minimum values (254.50 and 42.40 N/m) at $\theta$ = 90\textdegree. In contrast, Young’s modulus of PO-Ge reaches its maximum value at $\theta$ = 90\textdegree (27.28 N/m) and its minimum value (8.26 N/m) at $\theta$ = 0\textdegree, while PO-Sn reaches its maximum value (13.60 N/m) at $\theta$ = 90\textdegree and its minimum value (5.81 N/m) at $\theta$ = 45\textdegree. 

Regarding Poisson's ratio values, PO-IV sheets have proven to be exceptionally versatile materials. Based on the polar plot of Poisson's ratio, the most anisotropic structure in this category is the PO-Sn [Fig.\ref{fig:moduli} (d)], with a maximum Poisson's ratio of 0.70, which is 7 times greater than the minimum value of 0.10. Next, PO-Ge [Fig.\ref{fig:moduli} (d)] shows a ratio of almost 3.5. Lastly, PO-C [Fig.\ref{fig:moduli} (a)] and PO-Si [fig.\ref{fig:moduli} (b)] are the less anisotropic structures, with ratio values of 1.53 and 1.36, respectively. Our results show that the maximum values of Poisson's ratio of PO-VI are higher than those of their respective hexagonal phases, except for PO-Ge, see Table~\ref{tab:constants}. Thus, we conclude that the anisotropy of the pentaoctite structures makes them potential candidates for designing new devices with tailored mechanical properties for advanced technological applications.



\subsection{Electronic properties}

\begin{figure*}[ht!]
\includegraphics[width=12cm,scale=0.08,clip, keepaspectratio]{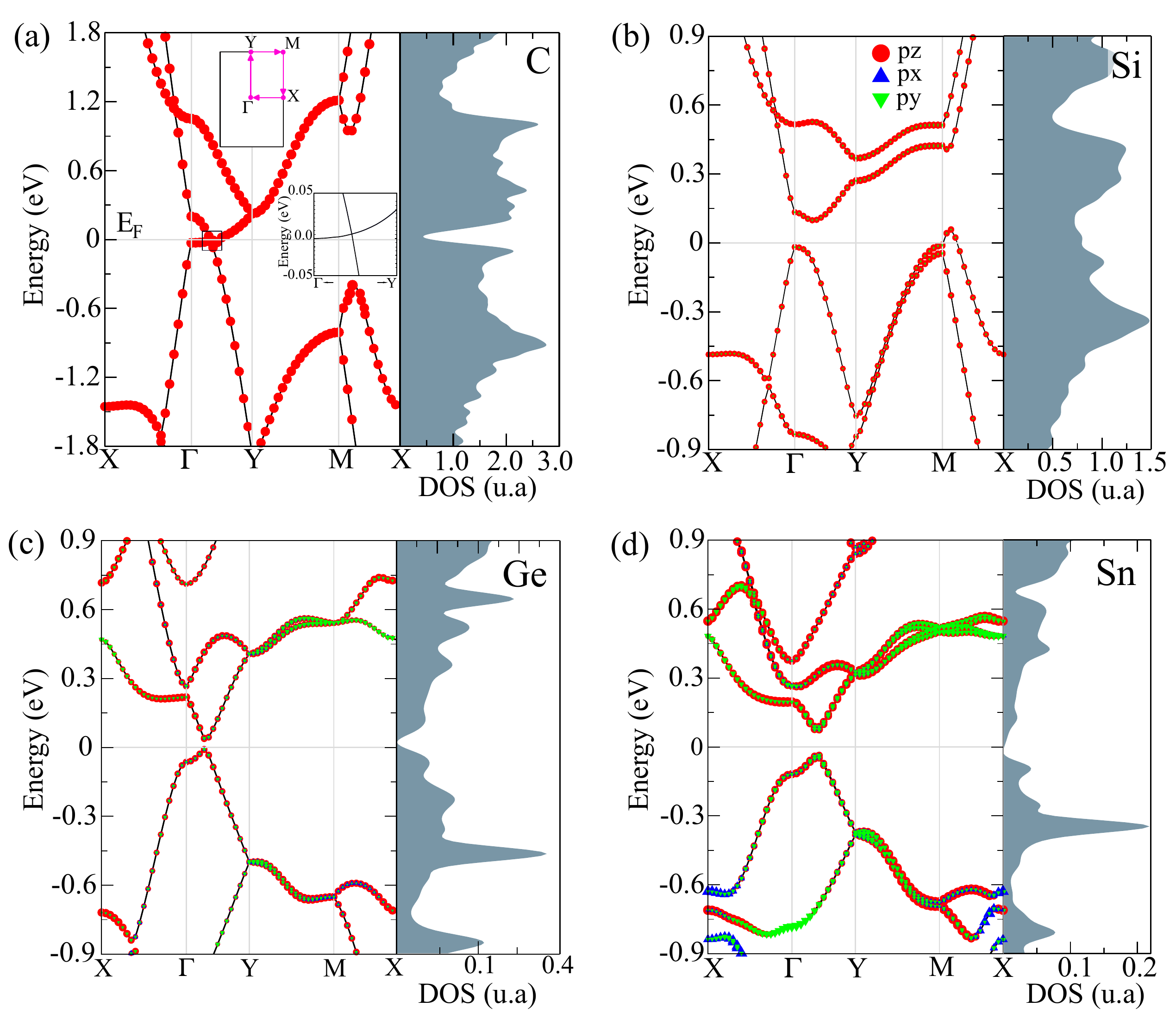}
\caption{\label{fig:bandas} Orbital-projected electronic band structures and density of states (DOS) of the (a) PO-C, (b) PO-Si, (c) PO-Ge, and (d) PO-Sn monolayers. The red (balls), blue(triangles up), and green(triangles down) symbols represent the s, p, and d orbitals contributions. The size of the symbols shows the relative contribution from the orbitals.} 
\end{figure*}

Fig.\ref{fig:bandas} shows the orbital-projected electronic band structures and density of states (DOS) of the PO-IV monolayers. The inset in Fig.\ref{fig:bandas}(a) illustrates the direction along the high symmetry points used to calculate the electronic band structure. PO-C exhibits a tilted Dirac cone at the Fermi level along the $\Gamma$$-$Y direction, as illustrated in Fig.\ref{fig:bandas}(a). Additionally, the DOS at the Fermi level provides evidence of the presence of a  Dirac cone. Unlike PO-C, 
which presents semimetallic features, PO-Si is a metal [Fig.\ref{fig:bandas}(b)], showing a hole pocket along the M$-$X direction. Furthermore, as shown in Figs.\ref{fig:bandas}(c) and \ref{fig:bandas}(d), PO-Ge and PO-Sn monolayers present a direct band gap along the $\Gamma$$-$Y direction, with band gap values of 45 and 111 meV, respectively.
Notably, the band gaps of PO-IV monolayers closely resemble those of their corresponding hexagonal phases: graphene (metal), silicene (1.55 meV), germanene (33 meV), and stanene (100 meV).\cite{novoselov2004,silicene-gap,germanene-gap,stanene-gap} Moreover, methods such as chemical functionalization, the application of external electric fields, and strain engineering offer additional pathways to optimize their optical and electronic behavior, positioning these materials at the forefront of 2D materials research.


\begin{figure}
    \centering
    \includegraphics[width=1.0\linewidth]{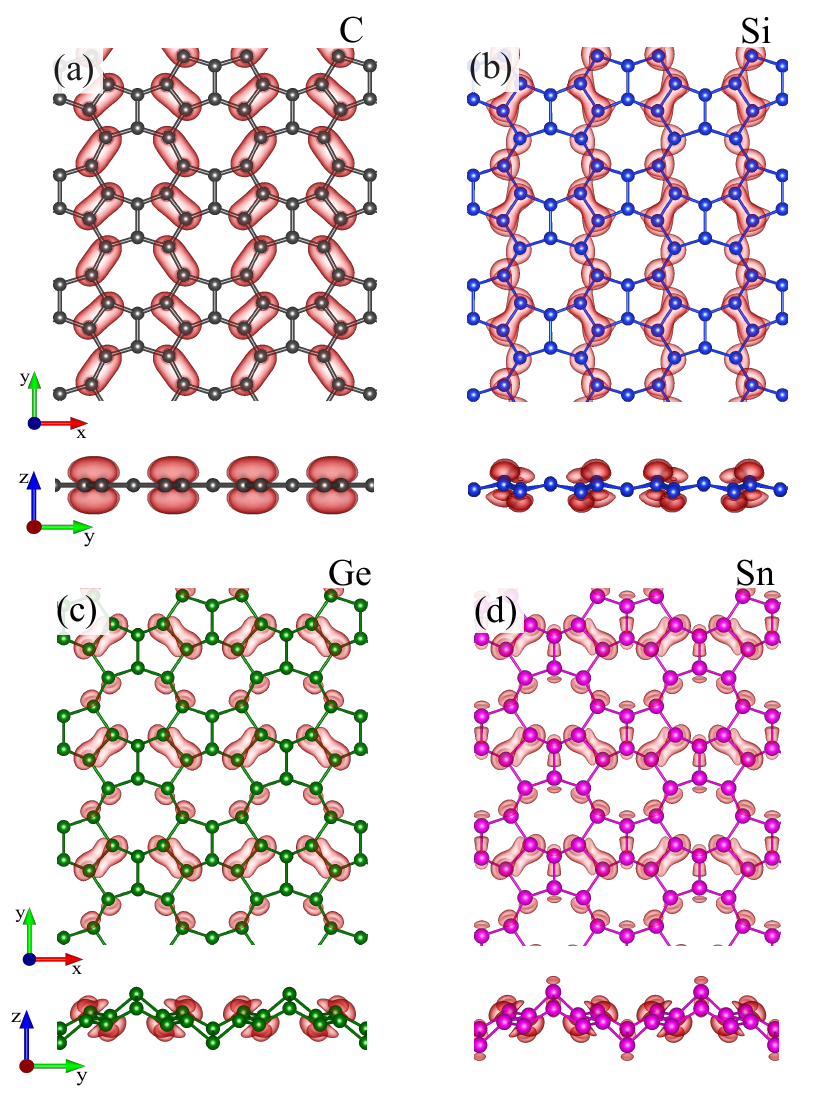}
    \caption{Top and front views of the partial charge density at $E - E_{F} \approx$ 100 meV of the (a) PO-C, (b) PO-Si, (c) PO-Ge, and (d) PO-Sn monolayers. Isosurface value is set 10$^{-4}$ e/\AA$^{3}$.}
    \label{fig:densitycharge}
\end{figure}

In general, the projections of the p$_z$ orbitals (red balls) dominate the states at the Fermi level for all studied PO-IV monolayers. For PO-C, the orbital projections indicate that most of the contribution comes from p$_z$ orbitals, while p$_x$ and p$_y$ orbitals exhibit more relevant contributions at deeper energy levels (not shown). The significant presence of p$_z$ states around the Fermi level is characteristic of 2D carbon-based systems like graphene and biphenylene. This prevalence of p$_z$ orbitals not only suggests the potential for constructing van der Waals heterostructures but also indicates reactivity favorable for adsorption processes, making these materials highly applicable in nanoelectronics and sensor devices. For the other structures, there is a noticeable reduction in the p$_z$ states, while the p$_y$ states become more prominent, as illustrated in Figs.\ref{fig:bandas}(b), \ref{fig:bandas}(c) and \ref{fig:bandas}(d).  

The influence of p$_z$ orbitals on the structural stability of single-layer materials is significant due to their role in $\pi$-bonding. For instance, in graphene, p$_z$ orbitals form a delocalized $\pi$-conjugated system, leading to a stable, flat structure.\cite{geim2007} However, in silicene, the larger and more diffuse p$_z$ orbitals of silicon atoms reduce the effective overlap and weaken $\pi$-conjugation. This leads to a buckled structure that minimizes energy through enhanced $\sigma$-bonding stabilization.\cite{Takahashi2017} 

To understand better the results observed in the orbital projections depicted at the band structures, we also analyzed the charge density of the lowest unoccupied states around $E - E_{F} \approx$ 100 meV, shown in Fig.\ref{fig:densitycharge}. In the PO-C sheet [Fig.\ref{fig:densitycharge}(a)], there is an extensive overlap of p$_z$ orbitals similar to the $\pi$-delocalized states observed in graphene. Meanwhile, Si, Ge, and Sn monolayers depicted in Figs.\ref{fig:densitycharge}(b),\ref{fig:densitycharge}(c) and \ref{fig:densitycharge}(d) show a less effective and more diffuse overlap of p$_z$ orbitals, similar to what is observed in silicene, germanene, and stanene.\cite{Takahashi2017,Chegel2020,stanene-gap}

The predominance of p$_z$ orbitals in the PO-C monolayer suggests a structural trend similar to graphene, resulting in a perfectly flat structure, as depicted in Fig.\ref{fig:estruturas} (a). On the other hand, the larger covalent radii of Si (1.11 \AA), Ge (1.20 \AA), and Sn (1.39 \AA) atoms compared to C (0.76 \AA) atom directly influence the p$_z$ orbital overlap, leading to reduced $\pi$-conjugation and a natural preference for a buckled structure. Furthermore, the effective overlap of p$_y$ and p$_z$ orbitals in these materials strongly enhances the structural buckling and stabilizes the system. As a result, in comparison to germanene\cite{Chegel2020}, which exhibits a mix of $sp^{2}$ and $sp^{3}$ hybridization within its buckled structure, our studied Si, Ge, and Sn sheets also display both $sp^{2}$ and $sp^{3}$ hybridization bonding. This behavior underscores the importance of p$_z$ orbitals in influencing the structural and electronic characteristics of PO-IV structures.

\subsection{PO-C HER performance}

\begin{figure*}[!htb]
\includegraphics[width=18cm,scale=0.5,clip, keepaspectratio]{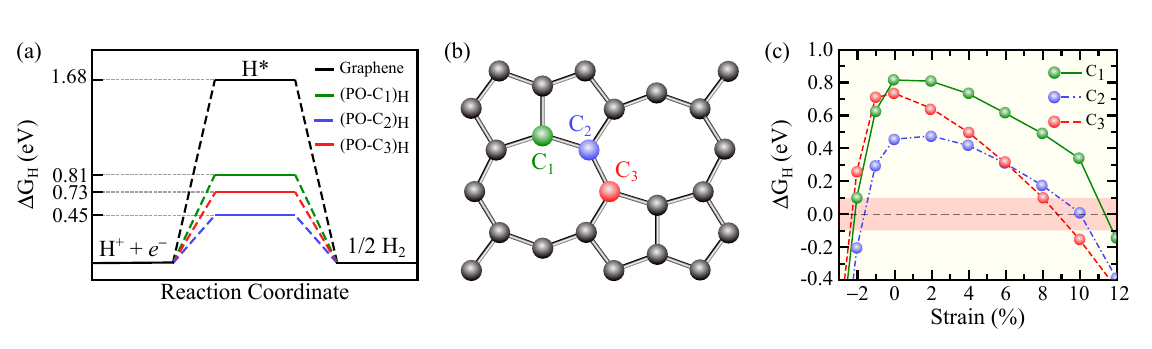}
\caption{\label{fig:HER} (a) Gibbs free energy ($\Delta G_{\rm H*}$) for hydrogen adsorption at the C$_1$, C$_2$, C$_3$ sites, represented in (b). $\Delta G_{\rm H*}$ for graphene is also included.  (c) $\Delta G_{\rm H*}$ values of the C$_1$, C$_2$, and C$_3$ configurations under biaxial strain conditions. The red region in (c) indicates values where $|\Delta G_{\rm H*}|\leq 0.1$~eV. The colors green, blue, and red refer to C$_1$, C$_2$, and C$_3$ sites, respectively. } 
\end{figure*}

As fossil fuel reserves decrease and the resulting environmental threats become more pressing, the development and use of clean, sustainable energy sources are rapidly accelerating. In this growing field of renewable options, hydrogen (H$_2$) emerges as a strong candidate for building a sustainable future.\cite{crabtree2008hydrogen} Recognized for its safety, absence of pollutants, effective utilization, and substantial energy yield, hydrogen carries significant promise in fueling our societal needs.  Exploring the viability of hydrogen storage has emerged as a prominent subject across the scientific community.\cite{abe2019hydrogen} Nevertheless, the inherently low intermolecular forces of H$_2$ molecules hinder their effective packing, creating challenges for storing H under ambient conditions.\cite{le2023fueling} A very well-explored route to produce H$_2$ is through the hydrogen evolution reaction (HER),\cite{lasia2019mechanism} which typically occurs at the surface of a metallic catalyst. 2D materials have been extensively investigated in the last few years as promising catalysts for HER.\cite{karmodak2020catalytic, li2021strategies} In this context, we investigated the catalytic activity of the metallic PO-C structure. 

To investigate the HER performance of the PO-C structure, we consider the systems in acidic media, where the reaction path of HER can be described
by the following steps:
\begin{eqnarray}
    H^+ + e^- + * &\rightarrow& H*; \label{volmer} \\
    H* + H^+ + e^- &\rightarrow& H_2 + *; \label{heyrovsky} \\
    2H* &\rightarrow& H_2 \label{tafel}.
\end{eqnarray}
Eq.~\ref{volmer} is a reduction of a proton ($H^+$) on an active site ($*$) of
the substrate (Volmer step), Eqs.~\ref{heyrovsky} and \ref{tafel} show the 
evolution of molecular $H_2$ through a second $H^+/e^-$ transfer (Heyrovsky step)
or by the combination of two adsorbed $H$ (Tafel step), respectively. 
The variation in the Gibbs free energy for the H adsorption ($\Delta G_{\rm H*}$) is
given by
\begin{equation}
    \label{eq:deltaG}
    \Delta G_{\rm H*} = \Delta E_{\rm H} + \Delta E_{\rm ZPE} - T\Delta S_{\rm H},
\end{equation}
where $\Delta S_{\rm H}$ represents the entropy change from before to after H adsorption at $T=298.15$~K, $\Delta E_{\rm ZPE}$ is the zero-point energy difference between adsorbed H and the gas phase (H$_2$), and 
$\Delta E_{\rm H} = E_{\rm sub+H} - E_{\rm sub} - \frac{1}{2}E_{\rm H_2}$ describes the adsorption energy of H$*$, with $E_{\rm sub+H}$, $E_{\rm sub}$ and H$_2$ represents the total energy of the substrate with adsorbed H, pristine substrate, and H$_2$ molecule, respectively. The $\Delta E_{\rm ZPE}$ and $T\Delta S_{\rm H}$ terms are nearly independent of the catalysts;\cite{norskov2005trends, zhou2019transition} therefore, we use the values described in Ref.~\citenum{norskov2005trends}, which
have been employed in other studies of HER on 2D materials.\cite{sahoo2023activation, qu2018effect, sajjad2023colossal, qu2015first, zhu2019single} Following this procedure, Eq.~\ref{eq:deltaG} can be written as $\Delta G_{\rm H*} = \Delta E_{\rm H} + 0.24$~eV.

In  Fig.~\ref{fig:HER}(a), we present $\Delta G_{\rm H*}$ values for hydrogenated PO-C at three different deposition sites, represented in Fig.~\ref{fig:HER}(b). 
For benchmark purposes, we also consider the graphene structure and obtain $\Delta G_{\rm H*}=1.68$~eV, which aligns closely with the 1.66~eV value reported in 
Ref.\cite{zhang2021improving}. With the definitions used, $\Delta G_{\rm H*} \rightarrow 0$ characterizes an ideal catalyst. Therefore, H adsorbed on the PO-C$_2$ site [(PO-C$_2$)$_{\rm H}$] exhibits the highest catalytic activity, with $\Delta G_{\rm H*} = 0.45$~eV.
The results indicate an energy difference of approximately 44\% for PO-C between C sites 1 and 2, with an intermediate value at C$_3$. Positive values suggest that the process of H adsorption onto the catalyst is energetically unfavorable from a kinetic standpoint; as a consequence, the PO-C structure has a higher catalytic efficiency than pristine graphene. Although it presents a lower catalytic property than biphenylene.\cite{bipheneleno-sci} 
It has been shown that in non-hexagonal rings, the strain caused by deviation from the ideal value of 120$^\circ$ in the $sp^2$ hybridization and antiaromaticity enhances the HER catalytic activity.\cite{liu2019crucial} Moreover, as it has been shown for other 2D structures, the PO-C performance might be enhanced through atomic defects or doping.\cite{cai2019design, tian2016enhanced,wang2019tunable} 


The HER performance can be tuned by applying strain, which has proven to be a highly effective strategy for altering the reactivity of metal catalysts.\cite{wang2019tunable} We show that applying strain to the PO-C structure allows for tuning its $\Delta G_{\rm H*}$ values, as presented in Fig.~\ref{fig:HER}(c). We observe a decrease in $\Delta G_{\rm H*}$ as the strain increases in all three analyzed cases. For (PO-C$_1$)$_{\rm H}$, (PO-C$_2$)$_{\rm H}$, and (PO-C$_3$)$_{\rm H}$ systems, at stretching levels of 11.34\%, 10\%, and 8.75\%, respectively, we find  $\Delta G_{\rm H*}$ $\rightarrow 0$~eV which is comparable to the catalytic properties of a Pt surface.\cite{norskov2005trends}. A compressive strain of -2\%, -1.54\%, and -2.25\% for the (PO-C$_1$)$_{\rm H}$, (PO-C$_2$)$_{\rm H}$, and (PO-C$_3$)$_{\rm H}$, respectively, also results in $\Delta G_{\rm H*}$ $\rightarrow 0$~eV. 
Notably, both the C$_2$ and C$3$ sites demonstrate high catalytic activity under tensile strains of 9.1\% to 9.54\%, with $\Delta G_{\rm H*}$ $<0.1$ eV, as shown in the red region of Fig.~\ref{fig:HER}(c). Similar activity is also observed for C$_{1}$ and C$_3$ sites under a compressive strain of -1.8\%. This indicates that, at specific strain values, the surface of the PO-C monolayer can have up to 66.6\% of their sites catalytically active for HER with $|\Delta G_{\rm H*}|\leq 0.1$ eV. 

Our findings reveal that strain engineering markedly enhances the hydrogen evolution reaction (HER) activity in PO-C sheets. Applying strain results in the creation of active sites with hydrogen adsorption energies ($|\Delta G|$) ranging from 0 to 0.1 eV, which are comparable to those of platinum (Pt)-based metallic catalysts.\cite{Pt-Co-HER,Pt-Ni-HER,Pt-Fe-HER} However, it is important to note that Pt-based catalysts suffer from issues like limited availability and poor stability.\cite{stability-Pt,stability-Pt-2} In addition, substituting Pt-based materials with alternative materials can significantly reduce the cost of the catalyst.\cite{stability-Pt-2} 
These findings demonstrate that strain engineering is a versatile and effective approach for optimizing the catalytic activity of the PO-C structure, offering a tunable method to enhance HER performance. Thus, our results strongly suggest that the PO-C monolayer holds great potential as a metal-free catalyst for HER.


\section{Conclusion}

In summary, we investigated the structural, electronic, and mechanical properties of the 2D allotrope of group-IV elements (C, Si, Ge, and Sn) called pentaoctite. Our findings reveal that in the $x$-direction, the PO-C monolayer is as rigid as graphene, while PO-Si, PO-Ge, and PO-Sn are more flexible than their hexagonal counterparts, which suggests a possible application for these sheets in foldable materials. Additionally, the 2D representation of Young’s modulus and Poisson’s ratio indicates that this new 2D allotrope exhibits highly anisotropic mechanical properties. 

The electronic band structure calculations reveal that PO-Ge and PO-Sn sheets are direct band gap semiconductors with values of 45 and 111 meV, while PO-Si and PO-C exhibit metallic properties with distinct electronic characteristics. Specifically, PO-Si is characterized by a valence band crossing the Fermi level, whereas PO-C features a tilted Dirac cone at the Fermi level.  The unique metallic properties of PO-C make it a promising candidate to be a catalyst for HER. The Gibbs free energy analysis demonstrates that under strain conditions, the PO-C sheet performs comparably to the best catalysts for HER found in the literature, e.g., the Pt-surface, with $\Delta G_{\rm H*} \rightarrow 0 $ eV.

Our findings indicate that the PO-IV monolayer is well-suited for nanoelectronics and nanomechanics, with PO-C, in particular, also serving as a metal-free electrocatalyst for HER. Additionally, PO-IV sheets show strong potential for experimental realization.

\begin{acknowledgments}

The authors acknowledge financial support from INCT-Materials Informatics and computer time from CENAPAD-UNICAMP and the Brazilian National Scientific Computing Laboratory (LNCC). I.S.S.d.O. acknowledges financial support from FAPEMIG (Project No. APQ-01425-21) and computer time from LCC-UFLA. T.A.S.P acknowledges PDPG-FAPDF-CAPES Centro-Oeste grant number 00193-00000867/2024-94 and support from CNPq grants 408144/2022-0 and 423423/2021-5. E.N.L. thanks to Augusto de Lelis Araújo for helpful discussions.

\end{acknowledgments}

\appendix

\bibliography{bib}

\begin{thebibliography}{73}%
\makeatletter
\providecommand \@ifxundefined [1]{%
 \@ifx{#1\undefined}
}%
\providecommand \@ifnum [1]{%
 \ifnum #1\expandafter \@firstoftwo
 \else \expandafter \@secondoftwo
 \fi
}%
\providecommand \@ifx [1]{%
 \ifx #1\expandafter \@firstoftwo
 \else \expandafter \@secondoftwo
 \fi
}%
\providecommand \natexlab [1]{#1}%
\providecommand \enquote  [1]{``#1''}%
\providecommand \bibnamefont  [1]{#1}%
\providecommand \bibfnamefont [1]{#1}%
\providecommand \citenamefont [1]{#1}%
\providecommand \href@noop [0]{\@secondoftwo}%
\providecommand \href [0]{\begingroup \@sanitize@url \@href}%
\providecommand \@href[1]{\@@startlink{#1}\@@href}%
\providecommand \@@href[1]{\endgroup#1\@@endlink}%
\providecommand \@sanitize@url [0]{\catcode `\\12\catcode `\$12\catcode
  `\&12\catcode `\#12\catcode `\^12\catcode `\_12\catcode `\%12\relax}%
\providecommand \@@startlink[1]{}%
\providecommand \@@endlink[0]{}%
\providecommand \url  [0]{\begingroup\@sanitize@url \@url }%
\providecommand \@url [1]{\endgroup\@href {#1}{\urlprefix }}%
\providecommand \urlprefix  [0]{URL }%
\providecommand \Eprint [0]{\href }%
\providecommand \doibase [0]{http://dx.doi.org/}%
\providecommand \selectlanguage [0]{\@gobble}%
\providecommand \bibinfo  [0]{\@secondoftwo}%
\providecommand \bibfield  [0]{\@secondoftwo}%
\providecommand \translation [1]{[#1]}%
\providecommand \BibitemOpen [0]{}%
\providecommand \bibitemStop [0]{}%
\providecommand \bibitemNoStop [0]{.\EOS\space}%
\providecommand \EOS [0]{\spacefactor3000\relax}%
\providecommand \BibitemShut  [1]{\csname bibitem#1\endcsname}%
\let\auto@bib@innerbib\@empty
\bibitem [{\citenamefont {Novoselov}\ \emph {et~al.}(2005)\citenamefont
  {Novoselov}, \citenamefont {D.~Jiang}, \citenamefont {Booth}, \citenamefont
  {Khotkevich}, \citenamefont {Morozov},\ and\ \citenamefont
  {Geim}}]{novoselov2005}%
  \BibitemOpen
  \bibfield  {author} {\bibinfo {author} {\bibfnamefont {K.~S.}\ \bibnamefont
  {Novoselov}}, \bibinfo {author} {\bibfnamefont {F.~Schedin}\ \bibnamefont
  {D.~Jiang}}, \bibinfo {author} {\bibfnamefont {T.~J.}\ \bibnamefont {Booth}},
  \bibinfo {author} {\bibfnamefont {V.~V.}\ \bibnamefont {Khotkevich}},
  \bibinfo {author} {\bibfnamefont {S.~V.}\ \bibnamefont {Morozov}}, \ and\
  \bibinfo {author} {\bibfnamefont {A.~K.}\ \bibnamefont {Geim}},\ }\bibfield
  {title} {\enquote {\bibinfo {title} {Two-dimensional atomic crystals},}\
  }\href {https://www.pnas.org/doi/abs/10.1073/pnas.0502848102} {\bibfield
  {journal} {\bibinfo  {journal} {Proceedings of the National Academy of
  Sciences}\ }\textbf {\bibinfo {volume} {102}},\ \bibinfo {pages} {10451}
  (\bibinfo {year} {2005})}\BibitemShut {NoStop}%
\bibitem [{\citenamefont {Novoselov}\ \emph {et~al.}(2004)\citenamefont
  {Novoselov}, \citenamefont {Geim}, \citenamefont {Morozov}, \citenamefont
  {Jiang}, \citenamefont {Zhang}, \citenamefont {Dubonos}, \citenamefont
  {Grigorieva},\ and\ \citenamefont {Firsov}}]{novoselov2004}%
  \BibitemOpen
  \bibfield  {author} {\bibinfo {author} {\bibfnamefont {K.~S.}\ \bibnamefont
  {Novoselov}}, \bibinfo {author} {\bibfnamefont {A.~K.}\ \bibnamefont {Geim}},
  \bibinfo {author} {\bibfnamefont {S.~V.}\ \bibnamefont {Morozov}}, \bibinfo
  {author} {\bibfnamefont {D.}~\bibnamefont {Jiang}}, \bibinfo {author}
  {\bibfnamefont {Y.}~\bibnamefont {Zhang}}, \bibinfo {author} {\bibfnamefont
  {S.~V.}\ \bibnamefont {Dubonos}}, \bibinfo {author} {\bibfnamefont {I.~V.}\
  \bibnamefont {Grigorieva}}, \ and\ \bibinfo {author} {\bibfnamefont {A.~A.}\
  \bibnamefont {Firsov}},\ }\bibfield  {title} {\enquote {\bibinfo {title}
  {Electric field effect in atomically thin carbon films},}\ }\href
  {https://www.science.org/doi/abs/10.1126/science.1102896} {\bibfield
  {journal} {\bibinfo  {journal} {Science}\ }\textbf {\bibinfo {volume}
  {306}},\ \bibinfo {pages} {666} (\bibinfo {year} {2004})}\BibitemShut
  {NoStop}%
\bibitem [{\citenamefont {Geim}\ and\ \citenamefont
  {Novoselov}(2007)}]{geim2007}%
  \BibitemOpen
  \bibfield  {author} {\bibinfo {author} {\bibfnamefont {A.~K.}\ \bibnamefont
  {Geim}}\ and\ \bibinfo {author} {\bibfnamefont {K.~S.}\ \bibnamefont
  {Novoselov}},\ }\bibfield  {title} {\enquote {\bibinfo {title} {The rise of
  graphene},}\ }\href {https://doi.org/10.1038/nmat18491} {\bibfield  {journal}
  {\bibinfo  {journal} {Nature Materials}\ }\textbf {\bibinfo {volume} {6}},\
  \bibinfo {pages} {183} (\bibinfo {year} {2007})}\BibitemShut {NoStop}%
\bibitem [{\citenamefont {Liu}\ and\ \citenamefont
  {Zhou}(2019)}]{liu2019recent}%
  \BibitemOpen
  \bibfield  {author} {\bibinfo {author} {\bibfnamefont {Bo}~\bibnamefont
  {Liu}}\ and\ \bibinfo {author} {\bibfnamefont {Kun}\ \bibnamefont {Zhou}},\
  }\bibfield  {title} {\enquote {\bibinfo {title} {Recent progress on
  graphene-analogous 2d nanomaterials: Properties, modeling and
  applications},}\ }\href
  {https://doi-org.ez26.periodicos.capes.gov.br/10.1016/j.pmatsci.2018.09.004}
  {\bibfield  {journal} {\bibinfo  {journal} {Progress in Materials Science}\
  }\textbf {\bibinfo {volume} {100}},\ \bibinfo {pages} {99--169} (\bibinfo
  {year} {2019})}\BibitemShut {NoStop}%
\bibitem [{\citenamefont {Cahangirov}\ \emph {et~al.}(2009)\citenamefont
  {Cahangirov}, \citenamefont {Topsakal}, \citenamefont {Akt{\"u}rk},
  \citenamefont {{\c{S}}ahin},\ and\ \citenamefont
  {Ciraci}}]{cahangirov2009two}%
  \BibitemOpen
  \bibfield  {author} {\bibinfo {author} {\bibfnamefont {Seymur}\ \bibnamefont
  {Cahangirov}}, \bibinfo {author} {\bibfnamefont {Mehmet}\ \bibnamefont
  {Topsakal}}, \bibinfo {author} {\bibfnamefont {Ethem}\ \bibnamefont
  {Akt{\"u}rk}}, \bibinfo {author} {\bibfnamefont {Hasan}\ \bibnamefont
  {{\c{S}}ahin}}, \ and\ \bibinfo {author} {\bibfnamefont {Salim}\ \bibnamefont
  {Ciraci}},\ }\bibfield  {title} {\enquote {\bibinfo {title} {Two-and
  one-dimensional honeycomb structures of silicon and germanium},}\ }\href@noop
  {} {\bibfield  {journal} {\bibinfo  {journal} {Physical review letters}\
  }\textbf {\bibinfo {volume} {102}},\ \bibinfo {pages} {236804} (\bibinfo
  {year} {2009})}\BibitemShut {NoStop}%
\bibitem [{\citenamefont {Liu}\ \emph {et~al.}(2011{\natexlab{a}})\citenamefont
  {Liu}, \citenamefont {Feng},\ and\ \citenamefont {Yao}}]{silicene-gap}%
  \BibitemOpen
  \bibfield  {author} {\bibinfo {author} {\bibfnamefont {Cheng-Cheng}\
  \bibnamefont {Liu}}, \bibinfo {author} {\bibfnamefont {Wanxiang}\
  \bibnamefont {Feng}}, \ and\ \bibinfo {author} {\bibfnamefont {Yugui}\
  \bibnamefont {Yao}},\ }\bibfield  {title} {\enquote {\bibinfo {title}
  {Quantum spin hall effect in silicene and two-dimensional germanium},}\
  }\href {https://doi.org/10.1103/PhysRevLett.107.076802} {\bibfield  {journal}
  {\bibinfo  {journal} {Physical review letters}\ }\textbf {\bibinfo {volume}
  {107}},\ \bibinfo {pages} {076802} (\bibinfo {year}
  {2011}{\natexlab{a}})}\BibitemShut {NoStop}%
\bibitem [{\citenamefont {Matthes}\ \emph {et~al.}(2013)\citenamefont
  {Matthes}, \citenamefont {Pulci},\ and\ \citenamefont
  {Bechstedt}}]{germanene-gap}%
  \BibitemOpen
  \bibfield  {author} {\bibinfo {author} {\bibfnamefont {Lars}\ \bibnamefont
  {Matthes}}, \bibinfo {author} {\bibfnamefont {Olivia}\ \bibnamefont {Pulci}},
  \ and\ \bibinfo {author} {\bibfnamefont {Friedhelm}\ \bibnamefont
  {Bechstedt}},\ }\bibfield  {title} {\enquote {\bibinfo {title} {Massive dirac
  quasiparticles in the optical absorbance of graphene, silicene, germanene,
  and tinene},}\ }\href {https://doi.org/10.1088/0953-8984/25/39/395305}
  {\bibfield  {journal} {\bibinfo  {journal} {Journal of Physics: Condensed
  Matter}\ }\textbf {\bibinfo {volume} {25}},\ \bibinfo {pages} {395305}
  (\bibinfo {year} {2013})}\BibitemShut {NoStop}%
\bibitem [{\citenamefont {Tang}\ \emph {et~al.}(2014)\citenamefont {Tang},
  \citenamefont {Chen}, \citenamefont {Cao}, \citenamefont {Huang},
  \citenamefont {Cahangirov}, \citenamefont {Xian}, \citenamefont {Xu},
  \citenamefont {Zhang}, \citenamefont {Duan},\ and\ \citenamefont
  {Rubio}}]{tang2014stable}%
  \BibitemOpen
  \bibfield  {author} {\bibinfo {author} {\bibfnamefont {Peizhe}\ \bibnamefont
  {Tang}}, \bibinfo {author} {\bibfnamefont {Pengcheng}\ \bibnamefont {Chen}},
  \bibinfo {author} {\bibfnamefont {Wendong}\ \bibnamefont {Cao}}, \bibinfo
  {author} {\bibfnamefont {Huaqing}\ \bibnamefont {Huang}}, \bibinfo {author}
  {\bibfnamefont {Seymur}\ \bibnamefont {Cahangirov}}, \bibinfo {author}
  {\bibfnamefont {Lede}\ \bibnamefont {Xian}}, \bibinfo {author} {\bibfnamefont
  {Yong}\ \bibnamefont {Xu}}, \bibinfo {author} {\bibfnamefont {Shou-Cheng}\
  \bibnamefont {Zhang}}, \bibinfo {author} {\bibfnamefont {Wenhui}\
  \bibnamefont {Duan}}, \ and\ \bibinfo {author} {\bibfnamefont {Angel}\
  \bibnamefont {Rubio}},\ }\bibfield  {title} {\enquote {\bibinfo {title}
  {Stable two-dimensional dumbbell stanene: A quantum spin hall insulator},}\
  }\href@noop {} {\bibfield  {journal} {\bibinfo  {journal} {Physical Review
  B}\ }\textbf {\bibinfo {volume} {90}},\ \bibinfo {pages} {121408} (\bibinfo
  {year} {2014})}\BibitemShut {NoStop}%
\bibitem [{\citenamefont {Zhu}\ \emph {et~al.}(2015{\natexlab{a}})\citenamefont
  {Zhu}, \citenamefont {Chen}, \citenamefont {Xu}, \citenamefont {Gao},
  \citenamefont {Guan}, \citenamefont {Liu}, \citenamefont {Qian},
  \citenamefont {Zhang},\ and\ \citenamefont {Jia}}]{zhu2015epitaxial}%
  \BibitemOpen
  \bibfield  {author} {\bibinfo {author} {\bibfnamefont {Feng-feng}\
  \bibnamefont {Zhu}}, \bibinfo {author} {\bibfnamefont {Wei-jiong}\
  \bibnamefont {Chen}}, \bibinfo {author} {\bibfnamefont {Yong}\ \bibnamefont
  {Xu}}, \bibinfo {author} {\bibfnamefont {Chun-lei}\ \bibnamefont {Gao}},
  \bibinfo {author} {\bibfnamefont {Dan-dan}\ \bibnamefont {Guan}}, \bibinfo
  {author} {\bibfnamefont {Can-hua}\ \bibnamefont {Liu}}, \bibinfo {author}
  {\bibfnamefont {Dong}\ \bibnamefont {Qian}}, \bibinfo {author} {\bibfnamefont
  {Shou-Cheng}\ \bibnamefont {Zhang}}, \ and\ \bibinfo {author} {\bibfnamefont
  {Jin-feng}\ \bibnamefont {Jia}},\ }\bibfield  {title} {\enquote {\bibinfo
  {title} {Epitaxial growth of two-dimensional stanene},}\ }\href@noop {}
  {\bibfield  {journal} {\bibinfo  {journal} {Nature materials}\ }\textbf
  {\bibinfo {volume} {14}},\ \bibinfo {pages} {1020--1025} (\bibinfo {year}
  {2015}{\natexlab{a}})}\BibitemShut {NoStop}%
\bibitem [{\citenamefont {Liu}\ \emph {et~al.}(2011{\natexlab{b}})\citenamefont
  {Liu}, \citenamefont {Feng},\ and\ \citenamefont {Yao}}]{liu2011quantum}%
  \BibitemOpen
  \bibfield  {author} {\bibinfo {author} {\bibfnamefont {Cheng-Cheng}\
  \bibnamefont {Liu}}, \bibinfo {author} {\bibfnamefont {Wanxiang}\
  \bibnamefont {Feng}}, \ and\ \bibinfo {author} {\bibfnamefont {Yugui}\
  \bibnamefont {Yao}},\ }\bibfield  {title} {\enquote {\bibinfo {title}
  {Quantum spin hall effect in silicene and two-dimensional germanium},}\
  }\href@noop {} {\bibfield  {journal} {\bibinfo  {journal} {Physical review
  letters}\ }\textbf {\bibinfo {volume} {107}},\ \bibinfo {pages} {076802}
  (\bibinfo {year} {2011}{\natexlab{b}})}\BibitemShut {NoStop}%
\bibitem [{\citenamefont {Van~den Broek}\ \emph {et~al.}(2014)\citenamefont
  {Van~den Broek}, \citenamefont {Houssa}, \citenamefont {Scalise},
  \citenamefont {Pourtois}, \citenamefont {Afanas‘ev},\ and\ \citenamefont
  {Stesmans}}]{van2014two}%
  \BibitemOpen
  \bibfield  {author} {\bibinfo {author} {\bibfnamefont {Bas}\ \bibnamefont
  {Van~den Broek}}, \bibinfo {author} {\bibfnamefont {Michel}\ \bibnamefont
  {Houssa}}, \bibinfo {author} {\bibfnamefont {Emilio}\ \bibnamefont
  {Scalise}}, \bibinfo {author} {\bibfnamefont {Geoffrey}\ \bibnamefont
  {Pourtois}}, \bibinfo {author} {\bibfnamefont {VV}~\bibnamefont
  {Afanas‘ev}}, \ and\ \bibinfo {author} {\bibfnamefont {Andre}\ \bibnamefont
  {Stesmans}},\ }\bibfield  {title} {\enquote {\bibinfo {title}
  {Two-dimensional hexagonal tin: ab initio geometry, stability, electronic
  structure and functionalization},}\ }\href@noop {} {\bibfield  {journal}
  {\bibinfo  {journal} {2D Materials}\ }\textbf {\bibinfo {volume} {1}},\
  \bibinfo {pages} {021004} (\bibinfo {year} {2014})}\BibitemShut {NoStop}%
\bibitem [{\citenamefont {Schwierz}(2010)}]{schwierz2010graphene}%
  \BibitemOpen
  \bibfield  {author} {\bibinfo {author} {\bibfnamefont {Frank}\ \bibnamefont
  {Schwierz}},\ }\bibfield  {title} {\enquote {\bibinfo {title} {Graphene
  transistors},}\ }\href@noop {} {\bibfield  {journal} {\bibinfo  {journal}
  {Nature nanotechnology}\ }\textbf {\bibinfo {volume} {5}},\ \bibinfo {pages}
  {487--496} (\bibinfo {year} {2010})}\BibitemShut {NoStop}%
\bibitem [{\citenamefont {Kang}\ \emph {et~al.}(2018)\citenamefont {Kang},
  \citenamefont {Wei},\ and\ \citenamefont {Li}}]{kang2018graphyne}%
  \BibitemOpen
  \bibfield  {author} {\bibinfo {author} {\bibfnamefont {Jun}\ \bibnamefont
  {Kang}}, \bibinfo {author} {\bibfnamefont {Zhongming}\ \bibnamefont {Wei}}, \
  and\ \bibinfo {author} {\bibfnamefont {Jingbo}\ \bibnamefont {Li}},\
  }\bibfield  {title} {\enquote {\bibinfo {title} {Graphyne and its family:
  recent theoretical advances},}\ }\href@noop {} {\bibfield  {journal}
  {\bibinfo  {journal} {ACS applied materials \& interfaces}\ }\textbf
  {\bibinfo {volume} {11}},\ \bibinfo {pages} {2692--2706} (\bibinfo {year}
  {2018})}\BibitemShut {NoStop}%
\bibitem [{\citenamefont {Li}\ \emph {et~al.}(2014)\citenamefont {Li},
  \citenamefont {Xu}, \citenamefont {Liu},\ and\ \citenamefont
  {Li}}]{li2014graphdiyne}%
  \BibitemOpen
  \bibfield  {author} {\bibinfo {author} {\bibfnamefont {Yongjun}\ \bibnamefont
  {Li}}, \bibinfo {author} {\bibfnamefont {Liang}\ \bibnamefont {Xu}}, \bibinfo
  {author} {\bibfnamefont {Huibiao}\ \bibnamefont {Liu}}, \ and\ \bibinfo
  {author} {\bibfnamefont {Yuliang}\ \bibnamefont {Li}},\ }\bibfield  {title}
  {\enquote {\bibinfo {title} {Graphdiyne and graphyne: from theoretical
  predictions to practical construction},}\ }\href@noop {} {\bibfield
  {journal} {\bibinfo  {journal} {Chemical Society Reviews}\ }\textbf {\bibinfo
  {volume} {43}},\ \bibinfo {pages} {2572--2586} (\bibinfo {year}
  {2014})}\BibitemShut {NoStop}%
\bibitem [{\citenamefont {Fan}\ \emph {et~al.}(2021{\natexlab{a}})\citenamefont
  {Fan}, \citenamefont {Yan}, \citenamefont {Tripp}, \citenamefont
  {Krej{\v{c}}{\'\i}}, \citenamefont {Dimosthenous}, \citenamefont {Kachel},
  \citenamefont {Chen}, \citenamefont {Foster}, \citenamefont {Koert},
  \citenamefont {Liljeroth} \emph {et~al.}}]{fan2021biphenylene}%
  \BibitemOpen
  \bibfield  {author} {\bibinfo {author} {\bibfnamefont {Qitang}\ \bibnamefont
  {Fan}}, \bibinfo {author} {\bibfnamefont {Linghao}\ \bibnamefont {Yan}},
  \bibinfo {author} {\bibfnamefont {Matthias~W}\ \bibnamefont {Tripp}},
  \bibinfo {author} {\bibfnamefont {Ond{\v{r}}ej}\ \bibnamefont
  {Krej{\v{c}}{\'\i}}}, \bibinfo {author} {\bibfnamefont {Stavrina}\
  \bibnamefont {Dimosthenous}}, \bibinfo {author} {\bibfnamefont {Stefan~R}\
  \bibnamefont {Kachel}}, \bibinfo {author} {\bibfnamefont {Mengyi}\
  \bibnamefont {Chen}}, \bibinfo {author} {\bibfnamefont {Adam~S}\ \bibnamefont
  {Foster}}, \bibinfo {author} {\bibfnamefont {Ulrich}\ \bibnamefont {Koert}},
  \bibinfo {author} {\bibfnamefont {Peter}\ \bibnamefont {Liljeroth}},  \emph
  {et~al.},\ }\bibfield  {title} {\enquote {\bibinfo {title} {Biphenylene
  network: A nonbenzenoid carbon allotrope},}\ }\href@noop {} {\bibfield
  {journal} {\bibinfo  {journal} {Science}\ }\textbf {\bibinfo {volume}
  {372}},\ \bibinfo {pages} {852--856} (\bibinfo {year}
  {2021}{\natexlab{a}})}\BibitemShut {NoStop}%
\bibitem [{\citenamefont {Enyashin}\ and\ \citenamefont
  {Ivanovskii}(2011)}]{enyashin2011graphene}%
  \BibitemOpen
  \bibfield  {author} {\bibinfo {author} {\bibfnamefont {Andrey~N}\
  \bibnamefont {Enyashin}}\ and\ \bibinfo {author} {\bibfnamefont
  {Alexander~L}\ \bibnamefont {Ivanovskii}},\ }\bibfield  {title} {\enquote
  {\bibinfo {title} {Graphene allotropes},}\ }\href@noop {} {\bibfield
  {journal} {\bibinfo  {journal} {physica status solidi (b)}\ }\textbf
  {\bibinfo {volume} {248}},\ \bibinfo {pages} {1879--1883} (\bibinfo {year}
  {2011})}\BibitemShut {NoStop}%
\bibitem [{\citenamefont {Jana}\ \emph {et~al.}(2021)\citenamefont {Jana},
  \citenamefont {Bandyopadhyay}, \citenamefont {Datta}, \citenamefont
  {Bhattacharya},\ and\ \citenamefont {Jana}}]{jana2021emerging}%
  \BibitemOpen
  \bibfield  {author} {\bibinfo {author} {\bibfnamefont {Susmita}\ \bibnamefont
  {Jana}}, \bibinfo {author} {\bibfnamefont {Arka}\ \bibnamefont
  {Bandyopadhyay}}, \bibinfo {author} {\bibfnamefont {Sujoy}\ \bibnamefont
  {Datta}}, \bibinfo {author} {\bibfnamefont {Debaprem}\ \bibnamefont
  {Bhattacharya}}, \ and\ \bibinfo {author} {\bibfnamefont {Debnarayan}\
  \bibnamefont {Jana}},\ }\bibfield  {title} {\enquote {\bibinfo {title}
  {Emerging properties of carbon based 2d material beyond graphene},}\
  }\href@noop {} {\bibfield  {journal} {\bibinfo  {journal} {Journal of
  Physics: Condensed Matter}\ }\textbf {\bibinfo {volume} {34}},\ \bibinfo
  {pages} {053001} (\bibinfo {year} {2021})}\BibitemShut {NoStop}%
\bibitem [{\citenamefont {Bollella}\ \emph {et~al.}(2017)\citenamefont
  {Bollella}, \citenamefont {Fusco}, \citenamefont {Tortolini}, \citenamefont
  {Sanz{\`o}}, \citenamefont {Favero}, \citenamefont {Gorton},\ and\
  \citenamefont {Antiochia}}]{bollella2017beyond}%
  \BibitemOpen
  \bibfield  {author} {\bibinfo {author} {\bibfnamefont {Paolo}\ \bibnamefont
  {Bollella}}, \bibinfo {author} {\bibfnamefont {Giovanni}\ \bibnamefont
  {Fusco}}, \bibinfo {author} {\bibfnamefont {Cristina}\ \bibnamefont
  {Tortolini}}, \bibinfo {author} {\bibfnamefont {Gabriella}\ \bibnamefont
  {Sanz{\`o}}}, \bibinfo {author} {\bibfnamefont {Gabriele}\ \bibnamefont
  {Favero}}, \bibinfo {author} {\bibfnamefont {Lo}~\bibnamefont {Gorton}}, \
  and\ \bibinfo {author} {\bibfnamefont {Riccarda}\ \bibnamefont {Antiochia}},\
  }\bibfield  {title} {\enquote {\bibinfo {title} {Beyond graphene:
  Electrochemical sensors and biosensors for biomarkers detection},}\
  }\href@noop {} {\bibfield  {journal} {\bibinfo  {journal} {Biosensors and
  Bioelectronics}\ }\textbf {\bibinfo {volume} {89}},\ \bibinfo {pages}
  {152--166} (\bibinfo {year} {2017})}\BibitemShut {NoStop}%
\bibitem [{\citenamefont {Peng}\ \emph {et~al.}(2014)\citenamefont {Peng},
  \citenamefont {Dearden}, \citenamefont {Crean}, \citenamefont {Han},
  \citenamefont {Liu}, \citenamefont {Wen},\ and\ \citenamefont
  {De}}]{peng2014new}%
  \BibitemOpen
  \bibfield  {author} {\bibinfo {author} {\bibfnamefont {Qing}\ \bibnamefont
  {Peng}}, \bibinfo {author} {\bibfnamefont {Albert~K}\ \bibnamefont
  {Dearden}}, \bibinfo {author} {\bibfnamefont {Jared}\ \bibnamefont {Crean}},
  \bibinfo {author} {\bibfnamefont {Liang}\ \bibnamefont {Han}}, \bibinfo
  {author} {\bibfnamefont {Sheng}\ \bibnamefont {Liu}}, \bibinfo {author}
  {\bibfnamefont {Xiaodong}\ \bibnamefont {Wen}}, \ and\ \bibinfo {author}
  {\bibfnamefont {Suvranu}\ \bibnamefont {De}},\ }\bibfield  {title} {\enquote
  {\bibinfo {title} {New materials graphyne, graphdiyne, graphone, and
  graphane: review of properties, synthesis, and application in
  nanotechnology},}\ }\href@noop {} {\bibfield  {journal} {\bibinfo  {journal}
  {Nanotechnology, science and applications}\ ,\ \bibinfo {pages} {1--29}}
  (\bibinfo {year} {2014})}\BibitemShut {NoStop}%
\bibitem [{\citenamefont {Zhang}\ \emph {et~al.}(2024)\citenamefont {Zhang},
  \citenamefont {Zhang}, \citenamefont {Mustafa}, \citenamefont {Saraswat},
  \citenamefont {Mekkey}, \citenamefont {Qassem}, \citenamefont {Karim},
  \citenamefont {Athab},\ and\ \citenamefont {Elmasry}}]{zhang2024application}%
  \BibitemOpen
  \bibfield  {author} {\bibinfo {author} {\bibfnamefont {Yuanyuan}\
  \bibnamefont {Zhang}}, \bibinfo {author} {\bibfnamefont {Zaizhen}\
  \bibnamefont {Zhang}}, \bibinfo {author} {\bibfnamefont {Mohammed~Ahmed}\
  \bibnamefont {Mustafa}}, \bibinfo {author} {\bibfnamefont {Shelesh~Krishna}\
  \bibnamefont {Saraswat}}, \bibinfo {author} {\bibfnamefont {Shereen~M}\
  \bibnamefont {Mekkey}}, \bibinfo {author} {\bibfnamefont {Laith~Yassen}\
  \bibnamefont {Qassem}}, \bibinfo {author} {\bibfnamefont {Manal~Morad}\
  \bibnamefont {Karim}}, \bibinfo {author} {\bibfnamefont {Ayat~H}\
  \bibnamefont {Athab}}, \ and\ \bibinfo {author} {\bibfnamefont {Yasser}\
  \bibnamefont {Elmasry}},\ }\bibfield  {title} {\enquote {\bibinfo {title}
  {Application of biphenylene nanosheets for metronidazole detection},}\
  }\href@noop {} {\bibfield  {journal} {\bibinfo  {journal} {Journal of
  Molecular Liquids}\ }\textbf {\bibinfo {volume} {398}},\ \bibinfo {pages}
  {124216} (\bibinfo {year} {2024})}\BibitemShut {NoStop}%
\bibitem [{\citenamefont {Lahiri}\ \emph {et~al.}(2010)\citenamefont {Lahiri},
  \citenamefont {Lin}, \citenamefont {Bozkurt}, \citenamefont {Oleynik},\ and\
  \citenamefont {Batzill}}]{lahiri2010extended}%
  \BibitemOpen
  \bibfield  {author} {\bibinfo {author} {\bibfnamefont {Jayeeta}\ \bibnamefont
  {Lahiri}}, \bibinfo {author} {\bibfnamefont {You}\ \bibnamefont {Lin}},
  \bibinfo {author} {\bibfnamefont {Pinar}\ \bibnamefont {Bozkurt}}, \bibinfo
  {author} {\bibfnamefont {Ivan~I}\ \bibnamefont {Oleynik}}, \ and\ \bibinfo
  {author} {\bibfnamefont {Matthias}\ \bibnamefont {Batzill}},\ }\bibfield
  {title} {\enquote {\bibinfo {title} {An extended defect in graphene as a
  metallic wire},}\ }\href@noop {} {\bibfield  {journal} {\bibinfo  {journal}
  {Nature nanotechnology}\ }\textbf {\bibinfo {volume} {5}},\ \bibinfo {pages}
  {326--329} (\bibinfo {year} {2010})}\BibitemShut {NoStop}%
\bibitem [{\citenamefont {Bhatt}\ \emph {et~al.}(2022)\citenamefont {Bhatt},
  \citenamefont {Kim},\ and\ \citenamefont {Kim}}]{bhatt2022various}%
  \BibitemOpen
  \bibfield  {author} {\bibinfo {author} {\bibfnamefont {Mahesh~Datt}\
  \bibnamefont {Bhatt}}, \bibinfo {author} {\bibfnamefont {Heeju}\ \bibnamefont
  {Kim}}, \ and\ \bibinfo {author} {\bibfnamefont {Gunn}\ \bibnamefont {Kim}},\
  }\bibfield  {title} {\enquote {\bibinfo {title} {Various defects in graphene:
  a review},}\ }\href@noop {} {\bibfield  {journal} {\bibinfo  {journal} {RSC
  advances}\ }\textbf {\bibinfo {volume} {12}},\ \bibinfo {pages}
  {21520--21547} (\bibinfo {year} {2022})}\BibitemShut {NoStop}%
\bibitem [{\citenamefont {Su}\ \emph {et~al.}(2013)\citenamefont {Su},
  \citenamefont {Jiang},\ and\ \citenamefont {Feng}}]{PO-C}%
  \BibitemOpen
  \bibfield  {author} {\bibinfo {author} {\bibfnamefont {Cong}\ \bibnamefont
  {Su}}, \bibinfo {author} {\bibfnamefont {Hua}\ \bibnamefont {Jiang}}, \ and\
  \bibinfo {author} {\bibfnamefont {Ji}~\bibnamefont {Feng}},\ }\bibfield
  {title} {\enquote {\bibinfo {title} {Two-dimensional carbon allotrope with
  strong electronic anisotropy},}\ }\href
  {https://doi.org/10.1103/PhysRevB.87.075453} {\bibfield  {journal} {\bibinfo
  {journal} {Physical Review B—Condensed Matter and Materials Physics}\
  }\textbf {\bibinfo {volume} {87}},\ \bibinfo {pages} {075453} (\bibinfo
  {year} {2013})}\BibitemShut {NoStop}%
\bibitem [{\citenamefont {Lima}\ \emph {et~al.}(2016)\citenamefont {Lima},
  \citenamefont {Schmidt},\ and\ \citenamefont
  {Nunes}}]{lima2016topologically}%
  \BibitemOpen
  \bibfield  {author} {\bibinfo {author} {\bibfnamefont {Erika~N}\ \bibnamefont
  {Lima}}, \bibinfo {author} {\bibfnamefont {Tome~M}\ \bibnamefont {Schmidt}},
  \ and\ \bibinfo {author} {\bibfnamefont {Ricardo~W}\ \bibnamefont {Nunes}},\
  }\bibfield  {title} {\enquote {\bibinfo {title} {Topologically protected
  metallic states induced by a one-dimensional extended defect in the bulk of a
  2d topological insulator},}\ }\href@noop {} {\bibfield  {journal} {\bibinfo
  {journal} {Nano Letters}\ }\textbf {\bibinfo {volume} {16}},\ \bibinfo
  {pages} {4025--4031} (\bibinfo {year} {2016})}\BibitemShut {NoStop}%
\bibitem [{\citenamefont {Lima}\ \emph {et~al.}(2019)\citenamefont {Lima},
  \citenamefont {Schmidt},\ and\ \citenamefont {Nunes}}]{erika1}%
  \BibitemOpen
  \bibfield  {author} {\bibinfo {author} {\bibfnamefont {Erika~N}\ \bibnamefont
  {Lima}}, \bibinfo {author} {\bibfnamefont {Tome~M}\ \bibnamefont {Schmidt}},
  \ and\ \bibinfo {author} {\bibfnamefont {R~W}\ \bibnamefont {Nunes}},\
  }\bibfield  {title} {\enquote {\bibinfo {title} {Structural and topological
  phase transitions induced by strain in two-dimensional bismuth},}\ }\href
  {\doibase 10.1088/1361-648X/ab3899} {\bibfield  {journal} {\bibinfo
  {journal} {Journal of Physics: Condensed Matter}\ }\textbf {\bibinfo {volume}
  {31}},\ \bibinfo {pages} {475001} (\bibinfo {year} {2019})}\BibitemShut
  {NoStop}%
\bibitem [{\citenamefont {da~Rosa}\ \emph {et~al.}(2021)\citenamefont
  {da~Rosa}, \citenamefont {Pontes}, \citenamefont {Lima},\ and\ \citenamefont
  {Frauenheim}}]{erika2}%
  \BibitemOpen
  \bibfield  {author} {\bibinfo {author} {\bibfnamefont {Andreia~Luisa}\
  \bibnamefont {da~Rosa}}, \bibinfo {author} {\bibfnamefont {Renato~Borges}\
  \bibnamefont {Pontes}}, \bibinfo {author} {\bibfnamefont {Erika~Nascimento}\
  \bibnamefont {Lima}}, \ and\ \bibinfo {author} {\bibfnamefont {Thomas}\
  \bibnamefont {Frauenheim}},\ }\bibfield  {title} {\enquote {\bibinfo {title}
  {New pentaoctite phase of group-v nanostructures},}\ }\href
  {https://doi.org/10.1002/pssb.202100112} {\bibfield  {journal} {\bibinfo
  {journal} {physica status solidi (b)}\ }\textbf {\bibinfo {volume} {258}},\
  \bibinfo {pages} {2100112} (\bibinfo {year} {2021})}\BibitemShut {NoStop}%
\bibitem [{\citenamefont {Hohenberg}\ and\ \citenamefont {Kohn}(1964)}]{DFT_1}%
  \BibitemOpen
  \bibfield  {author} {\bibinfo {author} {\bibfnamefont {P.}~\bibnamefont
  {Hohenberg}}\ and\ \bibinfo {author} {\bibfnamefont {W.}~\bibnamefont
  {Kohn}},\ }\bibfield  {title} {\enquote {\bibinfo {title} {Inhomogeneous
  electron gas},}\ }\href {\doibase 10.1103/PhysRev.136.B864} {\bibfield
  {journal} {\bibinfo  {journal} {Phys. Rev.}\ }\textbf {\bibinfo {volume}
  {136}},\ \bibinfo {pages} {B864--B871} (\bibinfo {year} {1964})}\BibitemShut
  {NoStop}%
\bibitem [{\citenamefont {Kohn}\ and\ \citenamefont {Sham}(1965)}]{DFT_2}%
  \BibitemOpen
  \bibfield  {author} {\bibinfo {author} {\bibfnamefont {W.}~\bibnamefont
  {Kohn}}\ and\ \bibinfo {author} {\bibfnamefont {L.~J.}\ \bibnamefont
  {Sham}},\ }\bibfield  {title} {\enquote {\bibinfo {title} {Self-consistent
  equations including exchange and correlation effects},}\ }\href {\doibase
  10.1103/PhysRev.140.A1133} {\bibfield  {journal} {\bibinfo  {journal} {Phys.
  Rev.}\ }\textbf {\bibinfo {volume} {140}},\ \bibinfo {pages} {A1133--A1138}
  (\bibinfo {year} {1965})}\BibitemShut {NoStop}%
\bibitem [{\citenamefont {Kresse}\ and\ \citenamefont
  {Furthm\"uller}(1996)}]{VASP}%
  \BibitemOpen
  \bibfield  {author} {\bibinfo {author} {\bibfnamefont {G.}~\bibnamefont
  {Kresse}}\ and\ \bibinfo {author} {\bibfnamefont {J.}~\bibnamefont
  {Furthm\"uller}},\ }\bibfield  {title} {\enquote {\bibinfo {title} {Efficient
  iterative schemes for ab initio total-energy calculations using a plane-wave
  basis set},}\ }\href {\doibase 10.1103/PhysRevB.54.11169} {\bibfield
  {journal} {\bibinfo  {journal} {Phys. Rev. B}\ }\textbf {\bibinfo {volume}
  {54}},\ \bibinfo {pages} {11169--11186} (\bibinfo {year} {1996})}\BibitemShut
  {NoStop}%
\bibitem [{\citenamefont {Perdew}\ \emph
  {et~al.}(1996{\natexlab{a}})\citenamefont {Perdew}, \citenamefont {Burke},\
  and\ \citenamefont {Ernzerhof}}]{GGA}%
  \BibitemOpen
  \bibfield  {author} {\bibinfo {author} {\bibfnamefont {John~P.}\ \bibnamefont
  {Perdew}}, \bibinfo {author} {\bibfnamefont {Kieron}\ \bibnamefont {Burke}},
  \ and\ \bibinfo {author} {\bibfnamefont {Matthias}\ \bibnamefont
  {Ernzerhof}},\ }\bibfield  {title} {\enquote {\bibinfo {title} {Generalized
  gradient approximation made simple},}\ }\href {\doibase
  10.1103/PhysRevLett.77.3865} {\bibfield  {journal} {\bibinfo  {journal}
  {Phys. Rev. Lett.}\ }\textbf {\bibinfo {volume} {77}},\ \bibinfo {pages}
  {3865--3868} (\bibinfo {year} {1996}{\natexlab{a}})}\BibitemShut {NoStop}%
\bibitem [{\citenamefont {Perdew}\ \emph
  {et~al.}(1996{\natexlab{b}})\citenamefont {Perdew}, \citenamefont {Burke},\
  and\ \citenamefont {Ernzerhof}}]{PAW_1}%
  \BibitemOpen
  \bibfield  {author} {\bibinfo {author} {\bibfnamefont {John~P.}\ \bibnamefont
  {Perdew}}, \bibinfo {author} {\bibfnamefont {Kieron}\ \bibnamefont {Burke}},
  \ and\ \bibinfo {author} {\bibfnamefont {Matthias}\ \bibnamefont
  {Ernzerhof}},\ }\bibfield  {title} {\enquote {\bibinfo {title} {Generalized
  gradient approximation made simple},}\ }\href {\doibase
  10.1103/PhysRevLett.77.3865} {\bibfield  {journal} {\bibinfo  {journal}
  {Phys. Rev. Lett.}\ }\textbf {\bibinfo {volume} {77}},\ \bibinfo {pages}
  {3865--3868} (\bibinfo {year} {1996}{\natexlab{b}})}\BibitemShut {NoStop}%
\bibitem [{\citenamefont {Kresse}\ and\ \citenamefont {Joubert}(1999)}]{PAW_2}%
  \BibitemOpen
  \bibfield  {author} {\bibinfo {author} {\bibfnamefont {G.}~\bibnamefont
  {Kresse}}\ and\ \bibinfo {author} {\bibfnamefont {D.}~\bibnamefont
  {Joubert}},\ }\bibfield  {title} {\enquote {\bibinfo {title} {From ultrasoft
  pseudopotentials to the projector augmented-wave method},}\ }\href {\doibase
  10.1103/PhysRevB.59.1758} {\bibfield  {journal} {\bibinfo  {journal} {Phys.
  Rev. B}\ }\textbf {\bibinfo {volume} {59}},\ \bibinfo {pages} {1758--1775}
  (\bibinfo {year} {1999})}\BibitemShut {NoStop}%
\bibitem [{\citenamefont {Monkhorst}\ and\ \citenamefont
  {Pack}(1976)}]{Monkhorst}%
  \BibitemOpen
  \bibfield  {author} {\bibinfo {author} {\bibfnamefont {Hendrik~J.}\
  \bibnamefont {Monkhorst}}\ and\ \bibinfo {author} {\bibfnamefont {James~D.}\
  \bibnamefont {Pack}},\ }\bibfield  {title} {\enquote {\bibinfo {title}
  {Special points for brillouin-zone integrations},}\ }\href {\doibase
  10.1103/PhysRevB.13.5188} {\bibfield  {journal} {\bibinfo  {journal} {Phys.
  Rev. B}\ }\textbf {\bibinfo {volume} {13}},\ \bibinfo {pages} {5188--5192}
  (\bibinfo {year} {1976})}\BibitemShut {NoStop}%
\bibitem [{\citenamefont {Togo}\ and\ \citenamefont {Tanaka}(2015)}]{Phonopy}%
  \BibitemOpen
  \bibfield  {author} {\bibinfo {author} {\bibfnamefont {Atsushi}\ \bibnamefont
  {Togo}}\ and\ \bibinfo {author} {\bibfnamefont {Isao}\ \bibnamefont
  {Tanaka}},\ }\bibfield  {title} {\enquote {\bibinfo {title} {First principles
  phonon calculations in materials science},}\ }\href {\doibase
  https://doi.org/10.1016/j.scriptamat.2015.07.021} {\bibfield  {journal}
  {\bibinfo  {journal} {Scripta Materialia}\ }\textbf {\bibinfo {volume}
  {108}},\ \bibinfo {pages} {1--5} (\bibinfo {year} {2015})}\BibitemShut
  {NoStop}%
\bibitem [{\citenamefont {Andersen}(1980)}]{andersen}%
  \BibitemOpen
  \bibfield  {author} {\bibinfo {author} {\bibfnamefont {Hans~C}\ \bibnamefont
  {Andersen}},\ }\bibfield  {title} {\enquote {\bibinfo {title} {Molecular
  dynamics simulations at constant pressure and/or temperature},}\ }\href@noop
  {} {\bibfield  {journal} {\bibinfo  {journal} {The Journal of chemical
  physics}\ }\textbf {\bibinfo {volume} {72}},\ \bibinfo {pages} {2384--2393}
  (\bibinfo {year} {1980})}\BibitemShut {NoStop}%
\bibitem [{\citenamefont {Wang}\ \emph {et~al.}(2021)\citenamefont {Wang},
  \citenamefont {Xu}, \citenamefont {Liu}, \citenamefont {Tang},\ and\
  \citenamefont {Geng}}]{VASPKIT}%
  \BibitemOpen
  \bibfield  {author} {\bibinfo {author} {\bibfnamefont {Vei}\ \bibnamefont
  {Wang}}, \bibinfo {author} {\bibfnamefont {Nan}\ \bibnamefont {Xu}}, \bibinfo
  {author} {\bibfnamefont {Jin-Cheng}\ \bibnamefont {Liu}}, \bibinfo {author}
  {\bibfnamefont {Gang}\ \bibnamefont {Tang}}, \ and\ \bibinfo {author}
  {\bibfnamefont {Wen-Tong}\ \bibnamefont {Geng}},\ }\bibfield  {title}
  {\enquote {\bibinfo {title} {Vaspkit: A user-friendly interface facilitating
  high-throughput computing and analysis using vasp code},}\ }\href@noop {}
  {\bibfield  {journal} {\bibinfo  {journal} {Computer Physics Communications}\
  }\textbf {\bibinfo {volume} {267}},\ \bibinfo {pages} {108033} (\bibinfo
  {year} {2021})}\BibitemShut {NoStop}%
\bibitem [{\citenamefont {Krukau}\ \emph {et~al.}(2006)\citenamefont {Krukau},
  \citenamefont {Vydrov}, \citenamefont {Izmaylov},\ and\ \citenamefont
  {Scuseria}}]{HSE06}%
  \BibitemOpen
  \bibfield  {author} {\bibinfo {author} {\bibfnamefont {Aliaksandr~V}\
  \bibnamefont {Krukau}}, \bibinfo {author} {\bibfnamefont {Oleg~A}\
  \bibnamefont {Vydrov}}, \bibinfo {author} {\bibfnamefont {Artur~F}\
  \bibnamefont {Izmaylov}}, \ and\ \bibinfo {author} {\bibfnamefont
  {Gustavo~E}\ \bibnamefont {Scuseria}},\ }\bibfield  {title} {\enquote
  {\bibinfo {title} {Influence of the exchange screening parameter on the
  performance of screened hybrid functionals},}\ }\href@noop {} {\bibfield
  {journal} {\bibinfo  {journal} {The Journal of chemical physics}\ }\textbf
  {\bibinfo {volume} {125}} (\bibinfo {year} {2006})}\BibitemShut {NoStop}%
\bibitem [{\citenamefont {Takeda}\ and\ \citenamefont
  {Shiraishi}(1994)}]{Si-Ge}%
  \BibitemOpen
  \bibfield  {author} {\bibinfo {author} {\bibfnamefont {Kyozaburo}\
  \bibnamefont {Takeda}}\ and\ \bibinfo {author} {\bibfnamefont {Kenji}\
  \bibnamefont {Shiraishi}},\ }\bibfield  {title} {\enquote {\bibinfo {title}
  {Theoretical possibility of stage corrugation in si and ge analogs of
  graphite},}\ }\href@noop {} {\bibfield  {journal} {\bibinfo  {journal}
  {Physical Review B}\ }\textbf {\bibinfo {volume} {50}},\ \bibinfo {pages}
  {14916} (\bibinfo {year} {1994})}\BibitemShut {NoStop}%
\bibitem [{\citenamefont {Zhu}\ \emph {et~al.}(2015{\natexlab{b}})\citenamefont
  {Zhu}, \citenamefont {Chen}, \citenamefont {Xu}, \citenamefont {Gao},
  \citenamefont {Guan}, \citenamefont {Liu}, \citenamefont {Qian},
  \citenamefont {Zhang},\ and\ \citenamefont {Jia}}]{Sn}%
  \BibitemOpen
  \bibfield  {author} {\bibinfo {author} {\bibfnamefont {Feng-feng}\
  \bibnamefont {Zhu}}, \bibinfo {author} {\bibfnamefont {Wei-jiong}\
  \bibnamefont {Chen}}, \bibinfo {author} {\bibfnamefont {Yong}\ \bibnamefont
  {Xu}}, \bibinfo {author} {\bibfnamefont {Chun-lei}\ \bibnamefont {Gao}},
  \bibinfo {author} {\bibfnamefont {Dan-dan}\ \bibnamefont {Guan}}, \bibinfo
  {author} {\bibfnamefont {Can-hua}\ \bibnamefont {Liu}}, \bibinfo {author}
  {\bibfnamefont {Dong}\ \bibnamefont {Qian}}, \bibinfo {author} {\bibfnamefont
  {Shou-Cheng}\ \bibnamefont {Zhang}}, \ and\ \bibinfo {author} {\bibfnamefont
  {Jin-feng}\ \bibnamefont {Jia}},\ }\bibfield  {title} {\enquote {\bibinfo
  {title} {Epitaxial growth of two-dimensional stanene},}\ }\href@noop {}
  {\bibfield  {journal} {\bibinfo  {journal} {Nature materials}\ }\textbf
  {\bibinfo {volume} {14}},\ \bibinfo {pages} {1020--1025} (\bibinfo {year}
  {2015}{\natexlab{b}})}\BibitemShut {NoStop}%
\bibitem [{\citenamefont {Fan}\ \emph {et~al.}(2021{\natexlab{b}})\citenamefont
  {Fan}, \citenamefont {Yan}, \citenamefont {Tripp}, \citenamefont {Krejčí},
  \citenamefont {Dimosthenous}, \citenamefont {Kachel}, \citenamefont {Chen},
  \citenamefont {Foster}, \citenamefont {Koert}, \citenamefont {Liljeroth},\
  and\ \citenamefont {Gottfried}}]{biphenele-sint}%
  \BibitemOpen
  \bibfield  {author} {\bibinfo {author} {\bibfnamefont {Qitang}\ \bibnamefont
  {Fan}}, \bibinfo {author} {\bibfnamefont {Linghao}\ \bibnamefont {Yan}},
  \bibinfo {author} {\bibfnamefont {Matthias~W.}\ \bibnamefont {Tripp}},
  \bibinfo {author} {\bibfnamefont {Ondřej}\ \bibnamefont {Krejčí}},
  \bibinfo {author} {\bibfnamefont {Stavrina}\ \bibnamefont {Dimosthenous}},
  \bibinfo {author} {\bibfnamefont {Stefan~R.}\ \bibnamefont {Kachel}},
  \bibinfo {author} {\bibfnamefont {Mengyi}\ \bibnamefont {Chen}}, \bibinfo
  {author} {\bibfnamefont {Adam~S.}\ \bibnamefont {Foster}}, \bibinfo {author}
  {\bibfnamefont {Ulrich}\ \bibnamefont {Koert}}, \bibinfo {author}
  {\bibfnamefont {Peter}\ \bibnamefont {Liljeroth}}, \ and\ \bibinfo {author}
  {\bibfnamefont {J.~Michael}\ \bibnamefont {Gottfried}},\ }\bibfield  {title}
  {\enquote {\bibinfo {title} {Biphenylene network: A nonbenzenoid carbon
  allotrope},}\ }\href {\doibase 10.1126/science.abg4509} {\bibfield  {journal}
  {\bibinfo  {journal} {Science}\ }\textbf {\bibinfo {volume} {372}},\ \bibinfo
  {pages} {852--856} (\bibinfo {year} {2021}{\natexlab{b}})}\BibitemShut
  {NoStop}%
\bibitem [{\citenamefont {Born}(1940)}]{born1}%
  \BibitemOpen
  \bibfield  {author} {\bibinfo {author} {\bibfnamefont {Max}\ \bibnamefont
  {Born}},\ }\bibfield  {title} {\enquote {\bibinfo {title} {On the stability
  of crystal lattices. i},}\ }in\ \href@noop {} {\emph {\bibinfo {booktitle}
  {Mathematical Proceedings of the Cambridge Philosophical Society}}},\
  Vol.~\bibinfo {volume} {36}\ (\bibinfo {organization} {Cambridge University
  Press},\ \bibinfo {year} {1940})\ pp.\ \bibinfo {pages}
  {160--172}\BibitemShut {NoStop}%
\bibitem [{\citenamefont {Born}\ and\ \citenamefont {Huang}(1996)}]{born2}%
  \BibitemOpen
  \bibfield  {author} {\bibinfo {author} {\bibfnamefont {Max}\ \bibnamefont
  {Born}}\ and\ \bibinfo {author} {\bibfnamefont {Kun}\ \bibnamefont {Huang}},\
  }\href@noop {} {\emph {\bibinfo {title} {Dynamical theory of crystal
  lattices}}}\ (\bibinfo  {publisher} {Oxford university press},\ \bibinfo
  {year} {1996})\BibitemShut {NoStop}%
\bibitem [{\citenamefont {Liao}\ \emph {et~al.}(2021)\citenamefont {Liao},
  \citenamefont {Shi}, \citenamefont {Ouyang}, \citenamefont {Li},
  \citenamefont {Zhang}, \citenamefont {Tang}, \citenamefont {He},\ and\
  \citenamefont {Zhong}}]{constant-C}%
  \BibitemOpen
  \bibfield  {author} {\bibinfo {author} {\bibfnamefont {Yujie}\ \bibnamefont
  {Liao}}, \bibinfo {author} {\bibfnamefont {XiZhi}\ \bibnamefont {Shi}},
  \bibinfo {author} {\bibfnamefont {Tao}\ \bibnamefont {Ouyang}}, \bibinfo
  {author} {\bibfnamefont {Jin}\ \bibnamefont {Li}}, \bibinfo {author}
  {\bibfnamefont {Chunxiao}\ \bibnamefont {Zhang}}, \bibinfo {author}
  {\bibfnamefont {Chao}\ \bibnamefont {Tang}}, \bibinfo {author} {\bibfnamefont
  {Chaoyu}\ \bibnamefont {He}}, \ and\ \bibinfo {author} {\bibfnamefont
  {Jianxin}\ \bibnamefont {Zhong}},\ }\bibfield  {title} {\enquote {\bibinfo
  {title} {New two-dimensional wide band gap hydrocarbon insulator by
  hydrogenation of a biphenylene sheet},}\ }\href@noop {} {\bibfield  {journal}
  {\bibinfo  {journal} {The Journal of Physical Chemistry Letters}\ }\textbf
  {\bibinfo {volume} {12}},\ \bibinfo {pages} {8889--8896} (\bibinfo {year}
  {2021})}\BibitemShut {NoStop}%
\bibitem [{\citenamefont {John}\ and\ \citenamefont
  {Merlin}(2016)}]{constants-Ge-Sn}%
  \BibitemOpen
  \bibfield  {author} {\bibinfo {author} {\bibfnamefont {Rita}\ \bibnamefont
  {John}}\ and\ \bibinfo {author} {\bibfnamefont {Benita}\ \bibnamefont
  {Merlin}},\ }\bibfield  {title} {\enquote {\bibinfo {title} {Theoretical
  investigation of structural, electronic, and mechanical properties of two
  dimensional c, si, ge, sn},}\ }\href@noop {} {\bibfield  {journal} {\bibinfo
  {journal} {Crystal Structure Theory and Applications}\ }\textbf {\bibinfo
  {volume} {5}},\ \bibinfo {pages} {43--55} (\bibinfo {year}
  {2016})}\BibitemShut {NoStop}%
\bibitem [{\citenamefont {Luo}\ \emph {et~al.}(2021)\citenamefont {Luo},
  \citenamefont {Ren}, \citenamefont {Xu}, \citenamefont {Yu}, \citenamefont
  {Wang},\ and\ \citenamefont {Sun}}]{bipheneleno-sci}%
  \BibitemOpen
  \bibfield  {author} {\bibinfo {author} {\bibfnamefont {Yi}~\bibnamefont
  {Luo}}, \bibinfo {author} {\bibfnamefont {Chongdan}\ \bibnamefont {Ren}},
  \bibinfo {author} {\bibfnamefont {Yujing}\ \bibnamefont {Xu}}, \bibinfo
  {author} {\bibfnamefont {Jin}\ \bibnamefont {Yu}}, \bibinfo {author}
  {\bibfnamefont {Sake}\ \bibnamefont {Wang}}, \ and\ \bibinfo {author}
  {\bibfnamefont {Minglei}\ \bibnamefont {Sun}},\ }\bibfield  {title} {\enquote
  {\bibinfo {title} {A first principles investigation on the structural,
  mechanical, electronic, and catalytic properties of biphenylene},}\ }\href
  {https://doi.org/10.1038/s41598-021-98261-9} {\bibfield  {journal} {\bibinfo
  {journal} {Scientific reports}\ }\textbf {\bibinfo {volume} {11}},\ \bibinfo
  {pages} {19008} (\bibinfo {year} {2021})}\BibitemShut {NoStop}%
\bibitem [{\citenamefont {Peng}\ and\ \citenamefont {De}(2013)}]{MoS2}%
  \BibitemOpen
  \bibfield  {author} {\bibinfo {author} {\bibfnamefont {Qing}\ \bibnamefont
  {Peng}}\ and\ \bibinfo {author} {\bibfnamefont {Suvranu}\ \bibnamefont
  {De}},\ }\bibfield  {title} {\enquote {\bibinfo {title} {Outstanding
  mechanical properties of monolayer mos 2 and its application in elastic
  energy storage},}\ }\href@noop {} {\bibfield  {journal} {\bibinfo  {journal}
  {Physical Chemistry Chemical Physics}\ }\textbf {\bibinfo {volume} {15}},\
  \bibinfo {pages} {19427--19437} (\bibinfo {year} {2013})}\BibitemShut
  {NoStop}%
\bibitem [{\citenamefont {Tao}\ \emph {et~al.}(2015)\citenamefont {Tao},
  \citenamefont {Shen}, \citenamefont {Wu}, \citenamefont {Liu}, \citenamefont
  {Feng}, \citenamefont {Wang}, \citenamefont {Hu}, \citenamefont {Yao},
  \citenamefont {Zhang}, \citenamefont {Pang} \emph
  {et~al.}}]{Black-phosphorus}%
  \BibitemOpen
  \bibfield  {author} {\bibinfo {author} {\bibfnamefont {Jin}\ \bibnamefont
  {Tao}}, \bibinfo {author} {\bibfnamefont {Wanfu}\ \bibnamefont {Shen}},
  \bibinfo {author} {\bibfnamefont {Sen}\ \bibnamefont {Wu}}, \bibinfo {author}
  {\bibfnamefont {Lu}~\bibnamefont {Liu}}, \bibinfo {author} {\bibfnamefont
  {Zhihong}\ \bibnamefont {Feng}}, \bibinfo {author} {\bibfnamefont {Chao}\
  \bibnamefont {Wang}}, \bibinfo {author} {\bibfnamefont {Chunguang}\
  \bibnamefont {Hu}}, \bibinfo {author} {\bibfnamefont {Pei}\ \bibnamefont
  {Yao}}, \bibinfo {author} {\bibfnamefont {Hao}\ \bibnamefont {Zhang}},
  \bibinfo {author} {\bibfnamefont {Wei}\ \bibnamefont {Pang}},  \emph
  {et~al.},\ }\bibfield  {title} {\enquote {\bibinfo {title} {Mechanical and
  electrical anisotropy of few-layer black phosphorus},}\ }\href@noop {}
  {\bibfield  {journal} {\bibinfo  {journal} {ACS nano}\ }\textbf {\bibinfo
  {volume} {9}},\ \bibinfo {pages} {11362--11370} (\bibinfo {year}
  {2015})}\BibitemShut {NoStop}%
\bibitem [{\citenamefont {Xu}\ \emph {et~al.}(2013)\citenamefont {Xu},
  \citenamefont {Yan}, \citenamefont {Zhang}, \citenamefont {Wang},
  \citenamefont {Xu}, \citenamefont {Tang}, \citenamefont {Duan},\ and\
  \citenamefont {Zhang}}]{stanene-gap}%
  \BibitemOpen
  \bibfield  {author} {\bibinfo {author} {\bibfnamefont {Yong}\ \bibnamefont
  {Xu}}, \bibinfo {author} {\bibfnamefont {Binghai}\ \bibnamefont {Yan}},
  \bibinfo {author} {\bibfnamefont {Hai-Jun}\ \bibnamefont {Zhang}}, \bibinfo
  {author} {\bibfnamefont {Jing}\ \bibnamefont {Wang}}, \bibinfo {author}
  {\bibfnamefont {Gang}\ \bibnamefont {Xu}}, \bibinfo {author} {\bibfnamefont
  {Peizhe}\ \bibnamefont {Tang}}, \bibinfo {author} {\bibfnamefont {Wenhui}\
  \bibnamefont {Duan}}, \ and\ \bibinfo {author} {\bibfnamefont {Shou-Cheng}\
  \bibnamefont {Zhang}},\ }\bibfield  {title} {\enquote {\bibinfo {title}
  {Large-gap quantum spin hall insulators in tin films},}\ }\href
  {https://doi.org/10.1103/PhysRevLett.111.136804} {\bibfield  {journal}
  {\bibinfo  {journal} {Physical review letters}\ }\textbf {\bibinfo {volume}
  {111}},\ \bibinfo {pages} {136804} (\bibinfo {year} {2013})}\BibitemShut
  {NoStop}%
\bibitem [{\citenamefont {Takahashi}(2017)}]{Takahashi2017}%
  \BibitemOpen
  \bibfield  {author} {\bibinfo {author} {\bibfnamefont {Masae}\ \bibnamefont
  {Takahashi}},\ }\bibfield  {title} {\enquote {\bibinfo {title} {Flat building
  blocks for flat silicene},}\ }\href {\doibase 10.1038/s41598-017-11360-4}
  {\bibfield  {journal} {\bibinfo  {journal} {Scientific Reports}\ }\textbf
  {\bibinfo {volume} {7}},\ \bibinfo {pages} {10855} (\bibinfo {year}
  {2017})}\BibitemShut {NoStop}%
\bibitem [{\citenamefont {Chegel}\ and\ \citenamefont
  {Behzad}(2020)}]{Chegel2020}%
  \BibitemOpen
  \bibfield  {author} {\bibinfo {author} {\bibfnamefont {Raad}\ \bibnamefont
  {Chegel}}\ and\ \bibinfo {author} {\bibfnamefont {Somayeh}\ \bibnamefont
  {Behzad}},\ }\bibfield  {title} {\enquote {\bibinfo {title} {Tunable
  electronic, optical, and thermal properties of two- dimensional germanene via
  an external electric field},}\ }\href {\doibase 10.1038/s41598-020-57558-x}
  {\bibfield  {journal} {\bibinfo  {journal} {Scientific Reports}\ }\textbf
  {\bibinfo {volume} {10}},\ \bibinfo {pages} {704} (\bibinfo {year}
  {2020})}\BibitemShut {NoStop}%
\bibitem [{\citenamefont {Crabtree}\ and\ \citenamefont
  {Dresselhaus}(2008)}]{crabtree2008hydrogen}%
  \BibitemOpen
  \bibfield  {author} {\bibinfo {author} {\bibfnamefont {George~W}\
  \bibnamefont {Crabtree}}\ and\ \bibinfo {author} {\bibfnamefont {Mildred~S}\
  \bibnamefont {Dresselhaus}},\ }\bibfield  {title} {\enquote {\bibinfo {title}
  {The hydrogen fuel alternative},}\ }\href@noop {} {\bibfield  {journal}
  {\bibinfo  {journal} {Mrs Bulletin}\ }\textbf {\bibinfo {volume} {33}},\
  \bibinfo {pages} {421--428} (\bibinfo {year} {2008})}\BibitemShut {NoStop}%
\bibitem [{\citenamefont {Abe}\ \emph {et~al.}(2019)\citenamefont {Abe},
  \citenamefont {Popoola}, \citenamefont {Ajenifuja},\ and\ \citenamefont
  {Popoola}}]{abe2019hydrogen}%
  \BibitemOpen
  \bibfield  {author} {\bibinfo {author} {\bibfnamefont {John~O}\ \bibnamefont
  {Abe}}, \bibinfo {author} {\bibfnamefont {API}\ \bibnamefont {Popoola}},
  \bibinfo {author} {\bibfnamefont {Emmanueal}\ \bibnamefont {Ajenifuja}}, \
  and\ \bibinfo {author} {\bibfnamefont {Olawale~M}\ \bibnamefont {Popoola}},\
  }\bibfield  {title} {\enquote {\bibinfo {title} {Hydrogen energy, economy and
  storage: Review and recommendation},}\ }\href@noop {} {\bibfield  {journal}
  {\bibinfo  {journal} {International journal of hydrogen energy}\ }\textbf
  {\bibinfo {volume} {44}},\ \bibinfo {pages} {15072--15086} (\bibinfo {year}
  {2019})}\BibitemShut {NoStop}%
\bibitem [{\citenamefont {Le}\ \emph {et~al.}(2023)\citenamefont {Le},
  \citenamefont {Sharma}, \citenamefont {Bora}, \citenamefont {Tran},
  \citenamefont {Truong}, \citenamefont {Le},\ and\ \citenamefont
  {Nguyen}}]{le2023fueling}%
  \BibitemOpen
  \bibfield  {author} {\bibinfo {author} {\bibfnamefont {Thanh~Tuan}\
  \bibnamefont {Le}}, \bibinfo {author} {\bibfnamefont {Prabhakar}\
  \bibnamefont {Sharma}}, \bibinfo {author} {\bibfnamefont {Bhaskor~Jyoti}\
  \bibnamefont {Bora}}, \bibinfo {author} {\bibfnamefont {Viet~Dung}\
  \bibnamefont {Tran}}, \bibinfo {author} {\bibfnamefont {Thanh~Hai}\
  \bibnamefont {Truong}}, \bibinfo {author} {\bibfnamefont {Huu~Cuong}\
  \bibnamefont {Le}}, \ and\ \bibinfo {author} {\bibfnamefont {Phuoc
  Quy~Phong}\ \bibnamefont {Nguyen}},\ }\bibfield  {title} {\enquote {\bibinfo
  {title} {Fueling the future: A comprehensive review of hydrogen energy
  systems and their challenges},}\ }\href@noop {} {\bibfield  {journal}
  {\bibinfo  {journal} {International Journal of Hydrogen Energy}\ } (\bibinfo
  {year} {2023})}\BibitemShut {NoStop}%
\bibitem [{\citenamefont {Lasia}(2019)}]{lasia2019mechanism}%
  \BibitemOpen
  \bibfield  {author} {\bibinfo {author} {\bibfnamefont {Andrzej}\ \bibnamefont
  {Lasia}},\ }\bibfield  {title} {\enquote {\bibinfo {title} {Mechanism and
  kinetics of the hydrogen evolution reaction},}\ }\href@noop {} {\bibfield
  {journal} {\bibinfo  {journal} {international journal of hydrogen energy}\
  }\textbf {\bibinfo {volume} {44}},\ \bibinfo {pages} {19484--19518} (\bibinfo
  {year} {2019})}\BibitemShut {NoStop}%
\bibitem [{\citenamefont {Karmodak}\ and\ \citenamefont
  {Andreussi}(2020)}]{karmodak2020catalytic}%
  \BibitemOpen
  \bibfield  {author} {\bibinfo {author} {\bibfnamefont {Naiwrit}\ \bibnamefont
  {Karmodak}}\ and\ \bibinfo {author} {\bibfnamefont {Oliviero}\ \bibnamefont
  {Andreussi}},\ }\bibfield  {title} {\enquote {\bibinfo {title} {Catalytic
  activity and stability of two-dimensional materials for the hydrogen
  evolution reaction},}\ }\href@noop {} {\bibfield  {journal} {\bibinfo
  {journal} {ACS Energy Letters}\ }\textbf {\bibinfo {volume} {5}},\ \bibinfo
  {pages} {885--891} (\bibinfo {year} {2020})}\BibitemShut {NoStop}%
\bibitem [{\citenamefont {Li}\ \emph {et~al.}(2021)\citenamefont {Li},
  \citenamefont {Sun},\ and\ \citenamefont {Guan}}]{li2021strategies}%
  \BibitemOpen
  \bibfield  {author} {\bibinfo {author} {\bibfnamefont {Saisai}\ \bibnamefont
  {Li}}, \bibinfo {author} {\bibfnamefont {Jianrui}\ \bibnamefont {Sun}}, \
  and\ \bibinfo {author} {\bibfnamefont {Jingqi}\ \bibnamefont {Guan}},\
  }\bibfield  {title} {\enquote {\bibinfo {title} {Strategies to improve
  electrocatalytic and photocatalytic performance of two-dimensional materials
  for hydrogen evolution reaction},}\ }\href@noop {} {\bibfield  {journal}
  {\bibinfo  {journal} {Chinese Journal of Catalysis}\ }\textbf {\bibinfo
  {volume} {42}},\ \bibinfo {pages} {511--556} (\bibinfo {year}
  {2021})}\BibitemShut {NoStop}%
\bibitem [{\citenamefont {N{\o}rskov}\ \emph {et~al.}(2005)\citenamefont
  {N{\o}rskov}, \citenamefont {Bligaard}, \citenamefont {Logadottir},
  \citenamefont {Kitchin}, \citenamefont {Chen}, \citenamefont {Pandelov},\
  and\ \citenamefont {Stimming}}]{norskov2005trends}%
  \BibitemOpen
  \bibfield  {author} {\bibinfo {author} {\bibfnamefont {Jens~Kehlet}\
  \bibnamefont {N{\o}rskov}}, \bibinfo {author} {\bibfnamefont {Thomas}\
  \bibnamefont {Bligaard}}, \bibinfo {author} {\bibfnamefont {Ashildur}\
  \bibnamefont {Logadottir}}, \bibinfo {author} {\bibfnamefont
  {JR}~\bibnamefont {Kitchin}}, \bibinfo {author} {\bibfnamefont {Jingguang~G}\
  \bibnamefont {Chen}}, \bibinfo {author} {\bibfnamefont {S}~\bibnamefont
  {Pandelov}}, \ and\ \bibinfo {author} {\bibfnamefont {U}~\bibnamefont
  {Stimming}},\ }\bibfield  {title} {\enquote {\bibinfo {title} {Trends in the
  exchange current for hydrogen evolution},}\ }\href@noop {} {\bibfield
  {journal} {\bibinfo  {journal} {Journal of The Electrochemical Society}\
  }\textbf {\bibinfo {volume} {152}},\ \bibinfo {pages} {J23} (\bibinfo {year}
  {2005})}\BibitemShut {NoStop}%
\bibitem [{\citenamefont {Zhou}\ \emph {et~al.}(2019)\citenamefont {Zhou},
  \citenamefont {Gao}, \citenamefont {Li}, \citenamefont {Chu},\ and\
  \citenamefont {Wang}}]{zhou2019transition}%
  \BibitemOpen
  \bibfield  {author} {\bibinfo {author} {\bibfnamefont {Yanan}\ \bibnamefont
  {Zhou}}, \bibinfo {author} {\bibfnamefont {Guoping}\ \bibnamefont {Gao}},
  \bibinfo {author} {\bibfnamefont {Yan}\ \bibnamefont {Li}}, \bibinfo {author}
  {\bibfnamefont {Wei}\ \bibnamefont {Chu}}, \ and\ \bibinfo {author}
  {\bibfnamefont {Lin-Wang}\ \bibnamefont {Wang}},\ }\bibfield  {title}
  {\enquote {\bibinfo {title} {Transition-metal single atoms in nitrogen-doped
  graphenes as efficient active centers for water splitting: a theoretical
  study},}\ }\href@noop {} {\bibfield  {journal} {\bibinfo  {journal} {Physical
  Chemistry Chemical Physics}\ }\textbf {\bibinfo {volume} {21}},\ \bibinfo
  {pages} {3024--3032} (\bibinfo {year} {2019})}\BibitemShut {NoStop}%
\bibitem [{\citenamefont {Sahoo}\ \emph {et~al.}(2023)\citenamefont {Sahoo},
  \citenamefont {Ray}, \citenamefont {Ahuja},\ and\ \citenamefont
  {Singh}}]{sahoo2023activation}%
  \BibitemOpen
  \bibfield  {author} {\bibinfo {author} {\bibfnamefont {Mihir~Ranjan}\
  \bibnamefont {Sahoo}}, \bibinfo {author} {\bibfnamefont {Avijeet}\
  \bibnamefont {Ray}}, \bibinfo {author} {\bibfnamefont {Rajeev}\ \bibnamefont
  {Ahuja}}, \ and\ \bibinfo {author} {\bibfnamefont {Nirpendra}\ \bibnamefont
  {Singh}},\ }\bibfield  {title} {\enquote {\bibinfo {title} {Activation of
  metal-free porous basal plane of biphenylene through defects engineering for
  hydrogen evolution reaction},}\ }\href@noop {} {\bibfield  {journal}
  {\bibinfo  {journal} {International Journal of Hydrogen Energy}\ }\textbf
  {\bibinfo {volume} {48}},\ \bibinfo {pages} {10545--10554} (\bibinfo {year}
  {2023})}\BibitemShut {NoStop}%
\bibitem [{\citenamefont {Qu}\ \emph {et~al.}(2018)\citenamefont {Qu},
  \citenamefont {Ke}, \citenamefont {Shao}, \citenamefont {Chen}, \citenamefont
  {Kwok}, \citenamefont {Shi},\ and\ \citenamefont {Pan}}]{qu2018effect}%
  \BibitemOpen
  \bibfield  {author} {\bibinfo {author} {\bibfnamefont {Yuanju}\ \bibnamefont
  {Qu}}, \bibinfo {author} {\bibfnamefont {Ye}~\bibnamefont {Ke}}, \bibinfo
  {author} {\bibfnamefont {Yangfan}\ \bibnamefont {Shao}}, \bibinfo {author}
  {\bibfnamefont {Wenzhou}\ \bibnamefont {Chen}}, \bibinfo {author}
  {\bibfnamefont {Chi~Tat}\ \bibnamefont {Kwok}}, \bibinfo {author}
  {\bibfnamefont {Xingqiang}\ \bibnamefont {Shi}}, \ and\ \bibinfo {author}
  {\bibfnamefont {Hui}\ \bibnamefont {Pan}},\ }\bibfield  {title} {\enquote
  {\bibinfo {title} {Effect of curvature on the hydrogen evolution reaction of
  graphene},}\ }\href@noop {} {\bibfield  {journal} {\bibinfo  {journal} {The
  Journal of Physical Chemistry C}\ }\textbf {\bibinfo {volume} {122}},\
  \bibinfo {pages} {25331--25338} (\bibinfo {year} {2018})}\BibitemShut
  {NoStop}%
\bibitem [{\citenamefont {Sajjad}\ \emph {et~al.}(2023)\citenamefont {Sajjad},
  \citenamefont {Nair}, \citenamefont {Samad},\ and\ \citenamefont
  {Singh}}]{sajjad2023colossal}%
  \BibitemOpen
  \bibfield  {author} {\bibinfo {author} {\bibfnamefont {Muhammad}\
  \bibnamefont {Sajjad}}, \bibinfo {author} {\bibfnamefont {Surabhi~Suresh}\
  \bibnamefont {Nair}}, \bibinfo {author} {\bibfnamefont {Yarjan~Abdul}\
  \bibnamefont {Samad}}, \ and\ \bibinfo {author} {\bibfnamefont {Nirpendra}\
  \bibnamefont {Singh}},\ }\bibfield  {title} {\enquote {\bibinfo {title}
  {Colossal figure of merit and compelling her catalytic activity of holey
  graphyne},}\ }\href@noop {} {\bibfield  {journal} {\bibinfo  {journal}
  {Scientific Reports}\ }\textbf {\bibinfo {volume} {13}},\ \bibinfo {pages}
  {9123} (\bibinfo {year} {2023})}\BibitemShut {NoStop}%
\bibitem [{\citenamefont {Qu}\ \emph {et~al.}(2015)\citenamefont {Qu},
  \citenamefont {Pan}, \citenamefont {Kwok},\ and\ \citenamefont
  {Wang}}]{qu2015first}%
  \BibitemOpen
  \bibfield  {author} {\bibinfo {author} {\bibfnamefont {Yuanju}\ \bibnamefont
  {Qu}}, \bibinfo {author} {\bibfnamefont {Hui}\ \bibnamefont {Pan}}, \bibinfo
  {author} {\bibfnamefont {Chi~Tat}\ \bibnamefont {Kwok}}, \ and\ \bibinfo
  {author} {\bibfnamefont {Zisheng}\ \bibnamefont {Wang}},\ }\bibfield  {title}
  {\enquote {\bibinfo {title} {A first-principles study on the hydrogen
  evolution reaction of vs 2 nanoribbons},}\ }\href@noop {} {\bibfield
  {journal} {\bibinfo  {journal} {Physical Chemistry Chemical Physics}\
  }\textbf {\bibinfo {volume} {17}},\ \bibinfo {pages} {24820--24825} (\bibinfo
  {year} {2015})}\BibitemShut {NoStop}%
\bibitem [{\citenamefont {Zhu}\ \emph {et~al.}(2019)\citenamefont {Zhu},
  \citenamefont {Hu}, \citenamefont {Wei},\ and\ \citenamefont
  {Hua}}]{zhu2019single}%
  \BibitemOpen
  \bibfield  {author} {\bibinfo {author} {\bibfnamefont {Hao-Ran}\ \bibnamefont
  {Zhu}}, \bibinfo {author} {\bibfnamefont {Yan-Ling}\ \bibnamefont {Hu}},
  \bibinfo {author} {\bibfnamefont {Shi-Hao}\ \bibnamefont {Wei}}, \ and\
  \bibinfo {author} {\bibfnamefont {Da-Yin}\ \bibnamefont {Hua}},\ }\bibfield
  {title} {\enquote {\bibinfo {title} {Single-metal atom anchored on boron
  monolayer ($\beta$12) as an electrocatalyst for nitrogen reduction into
  ammonia at ambient conditions: a first-principles study},}\ }\href@noop {}
  {\bibfield  {journal} {\bibinfo  {journal} {The Journal of Physical Chemistry
  C}\ }\textbf {\bibinfo {volume} {123}},\ \bibinfo {pages} {4274--4281}
  (\bibinfo {year} {2019})}\BibitemShut {NoStop}%
\bibitem [{\citenamefont {Zhang}\ \emph {et~al.}(2021)\citenamefont {Zhang},
  \citenamefont {Fu}, \citenamefont {Song},\ and\ \citenamefont
  {Wu}}]{zhang2021improving}%
  \BibitemOpen
  \bibfield  {author} {\bibinfo {author} {\bibfnamefont {Bo}~\bibnamefont
  {Zhang}}, \bibinfo {author} {\bibfnamefont {Xiuli}\ \bibnamefont {Fu}},
  \bibinfo {author} {\bibfnamefont {Li}~\bibnamefont {Song}}, \ and\ \bibinfo
  {author} {\bibfnamefont {Xiaojun}\ \bibnamefont {Wu}},\ }\bibfield  {title}
  {\enquote {\bibinfo {title} {Improving hydrogen evolution reaction
  performance by combining ditungsten carbide and nitrogen-doped graphene: A
  first-principles study},}\ }\href@noop {} {\bibfield  {journal} {\bibinfo
  {journal} {Carbon}\ }\textbf {\bibinfo {volume} {172}},\ \bibinfo {pages}
  {122--131} (\bibinfo {year} {2021})}\BibitemShut {NoStop}%
\bibitem [{\citenamefont {Liu}\ \emph {et~al.}(2019)\citenamefont {Liu},
  \citenamefont {Yu}, \citenamefont {Huang},\ and\ \citenamefont
  {Chen}}]{liu2019crucial}%
  \BibitemOpen
  \bibfield  {author} {\bibinfo {author} {\bibfnamefont {Jingwei}\ \bibnamefont
  {Liu}}, \bibinfo {author} {\bibfnamefont {Guangtao}\ \bibnamefont {Yu}},
  \bibinfo {author} {\bibfnamefont {Xuri}\ \bibnamefont {Huang}}, \ and\
  \bibinfo {author} {\bibfnamefont {Wei}\ \bibnamefont {Chen}},\ }\bibfield
  {title} {\enquote {\bibinfo {title} {The crucial role of strained ring in
  enhancing the hydrogen evolution catalytic activity for the 2d carbon
  allotropes: a high-throughput first-principles investigation},}\ }\href@noop
  {} {\bibfield  {journal} {\bibinfo  {journal} {2D Materials}\ }\textbf
  {\bibinfo {volume} {7}},\ \bibinfo {pages} {015015} (\bibinfo {year}
  {2019})}\BibitemShut {NoStop}%
\bibitem [{\citenamefont {Cai}\ \emph {et~al.}(2019)\citenamefont {Cai},
  \citenamefont {Gao}, \citenamefont {Chen}, \citenamefont {Ke}, \citenamefont
  {Zhang},\ and\ \citenamefont {Zhang}}]{cai2019design}%
  \BibitemOpen
  \bibfield  {author} {\bibinfo {author} {\bibfnamefont {Yongqing}\
  \bibnamefont {Cai}}, \bibinfo {author} {\bibfnamefont {Junfeng}\ \bibnamefont
  {Gao}}, \bibinfo {author} {\bibfnamefont {Shuai}\ \bibnamefont {Chen}},
  \bibinfo {author} {\bibfnamefont {Qingqing}\ \bibnamefont {Ke}}, \bibinfo
  {author} {\bibfnamefont {Gang}\ \bibnamefont {Zhang}}, \ and\ \bibinfo
  {author} {\bibfnamefont {Yong-Wei}\ \bibnamefont {Zhang}},\ }\bibfield
  {title} {\enquote {\bibinfo {title} {Design of phosphorene for hydrogen
  evolution performance comparable to platinum},}\ }\href@noop {} {\bibfield
  {journal} {\bibinfo  {journal} {Chemistry of Materials}\ }\textbf {\bibinfo
  {volume} {31}},\ \bibinfo {pages} {8948--8956} (\bibinfo {year}
  {2019})}\BibitemShut {NoStop}%
\bibitem [{\citenamefont {Tian}\ \emph {et~al.}(2016)\citenamefont {Tian},
  \citenamefont {Mei}, \citenamefont {Xue}, \citenamefont {Zhang},\ and\
  \citenamefont {Peng}}]{tian2016enhanced}%
  \BibitemOpen
  \bibfield  {author} {\bibinfo {author} {\bibfnamefont {Ye}~\bibnamefont
  {Tian}}, \bibinfo {author} {\bibfnamefont {Rui}\ \bibnamefont {Mei}},
  \bibinfo {author} {\bibfnamefont {Da-zhong}\ \bibnamefont {Xue}}, \bibinfo
  {author} {\bibfnamefont {Xiao}\ \bibnamefont {Zhang}}, \ and\ \bibinfo
  {author} {\bibfnamefont {Wei}\ \bibnamefont {Peng}},\ }\bibfield  {title}
  {\enquote {\bibinfo {title} {Enhanced electrocatalytic hydrogen evolution in
  graphene via defect engineering and heteroatoms co-doping},}\ }\href@noop {}
  {\bibfield  {journal} {\bibinfo  {journal} {Electrochimica Acta}\ }\textbf
  {\bibinfo {volume} {219}},\ \bibinfo {pages} {781--789} (\bibinfo {year}
  {2016})}\BibitemShut {NoStop}%
\bibitem [{\citenamefont {Wang}\ \emph {et~al.}(2019)\citenamefont {Wang},
  \citenamefont {Zeng}, \citenamefont {Gao}, \citenamefont {Maxson},
  \citenamefont {Raciti}, \citenamefont {Giroux}, \citenamefont {Pan},
  \citenamefont {Wang},\ and\ \citenamefont {Greeley}}]{wang2019tunable}%
  \BibitemOpen
  \bibfield  {author} {\bibinfo {author} {\bibfnamefont {Lei}\ \bibnamefont
  {Wang}}, \bibinfo {author} {\bibfnamefont {Zhenhua}\ \bibnamefont {Zeng}},
  \bibinfo {author} {\bibfnamefont {Wenpei}\ \bibnamefont {Gao}}, \bibinfo
  {author} {\bibfnamefont {Tristan}\ \bibnamefont {Maxson}}, \bibinfo {author}
  {\bibfnamefont {David}\ \bibnamefont {Raciti}}, \bibinfo {author}
  {\bibfnamefont {Michael}\ \bibnamefont {Giroux}}, \bibinfo {author}
  {\bibfnamefont {Xiaoqing}\ \bibnamefont {Pan}}, \bibinfo {author}
  {\bibfnamefont {Chao}\ \bibnamefont {Wang}}, \ and\ \bibinfo {author}
  {\bibfnamefont {Jeffrey}\ \bibnamefont {Greeley}},\ }\bibfield  {title}
  {\enquote {\bibinfo {title} {Tunable intrinsic strain in two-dimensional
  transition metal electrocatalysts},}\ }\href@noop {} {\bibfield  {journal}
  {\bibinfo  {journal} {Science}\ }\textbf {\bibinfo {volume} {363}},\ \bibinfo
  {pages} {870--874} (\bibinfo {year} {2019})}\BibitemShut {NoStop}%
\bibitem [{\citenamefont {Chen}\ \emph {et~al.}(2015)\citenamefont {Chen},
  \citenamefont {Su}, \citenamefont {Wang}, \citenamefont {Wu},\ and\
  \citenamefont {Zeng}}]{Pt-Co-HER}%
  \BibitemOpen
  \bibfield  {author} {\bibinfo {author} {\bibfnamefont {Sheng}\ \bibnamefont
  {Chen}}, \bibinfo {author} {\bibfnamefont {Hongyang}\ \bibnamefont {Su}},
  \bibinfo {author} {\bibfnamefont {Youcheng}\ \bibnamefont {Wang}}, \bibinfo
  {author} {\bibfnamefont {Wenlong}\ \bibnamefont {Wu}}, \ and\ \bibinfo
  {author} {\bibfnamefont {Jie}\ \bibnamefont {Zeng}},\ }\bibfield  {title}
  {\enquote {\bibinfo {title} {Size-controlled synthesis of platinum--copper
  hierarchical trigonal bipyramid nanoframes},}\ }\href
  {https://doi.org/10.1002/anie.201408399} {\bibfield  {journal} {\bibinfo
  {journal} {Angewandte Chemie}\ }\textbf {\bibinfo {volume} {127}},\ \bibinfo
  {pages} {110--115} (\bibinfo {year} {2015})}\BibitemShut {NoStop}%
\bibitem [{\citenamefont {Gu}\ \emph {et~al.}(2015)\citenamefont {Gu},
  \citenamefont {Lan}, \citenamefont {Jiang}, \citenamefont {Xu}, \citenamefont
  {Zhu}, \citenamefont {Jin},\ and\ \citenamefont {Zhang}}]{Pt-Ni-HER}%
  \BibitemOpen
  \bibfield  {author} {\bibinfo {author} {\bibfnamefont {Jun}\ \bibnamefont
  {Gu}}, \bibinfo {author} {\bibfnamefont {Guangxu}\ \bibnamefont {Lan}},
  \bibinfo {author} {\bibfnamefont {Yingying}\ \bibnamefont {Jiang}}, \bibinfo
  {author} {\bibfnamefont {Yanshuang}\ \bibnamefont {Xu}}, \bibinfo {author}
  {\bibfnamefont {Wei}\ \bibnamefont {Zhu}}, \bibinfo {author} {\bibfnamefont
  {Chuanhong}\ \bibnamefont {Jin}}, \ and\ \bibinfo {author} {\bibfnamefont
  {Yawen}\ \bibnamefont {Zhang}},\ }\bibfield  {title} {\enquote {\bibinfo
  {title} {Shaped pt-ni nanocrystals with an ultrathin pt-enriched shell
  derived from one-pot hydrothermal synthesis as active electrocatalysts for
  oxygen reduction},}\ }\href@noop {} {\bibfield  {journal} {\bibinfo
  {journal} {Nano Research}\ }\textbf {\bibinfo {volume} {8}},\ \bibinfo
  {pages} {1480--1496} (\bibinfo {year} {2015})}\BibitemShut {NoStop}%
\bibitem [{\citenamefont {Lee}\ \emph {et~al.}(2017)\citenamefont {Lee},
  \citenamefont {Jang}, \citenamefont {Xu}, \citenamefont {Feng}, \citenamefont
  {Park}, \citenamefont {Kim},\ and\ \citenamefont {Kim}}]{Pt-Fe-HER}%
  \BibitemOpen
  \bibfield  {author} {\bibinfo {author} {\bibfnamefont {Ji-Eun}\ \bibnamefont
  {Lee}}, \bibinfo {author} {\bibfnamefont {Yu~Jin}\ \bibnamefont {Jang}},
  \bibinfo {author} {\bibfnamefont {Wenqian}\ \bibnamefont {Xu}}, \bibinfo
  {author} {\bibfnamefont {Zhenxing}\ \bibnamefont {Feng}}, \bibinfo {author}
  {\bibfnamefont {Hee-Young}\ \bibnamefont {Park}}, \bibinfo {author}
  {\bibfnamefont {Jin~Young}\ \bibnamefont {Kim}}, \ and\ \bibinfo {author}
  {\bibfnamefont {Dong~Ha}\ \bibnamefont {Kim}},\ }\bibfield  {title} {\enquote
  {\bibinfo {title} {Ptfe nanoparticles supported on electroactive au--pani
  core@ shell nanoparticles for high performance bifunctional
  electrocatalysis},}\ }\href@noop {} {\bibfield  {journal} {\bibinfo
  {journal} {Journal of Materials Chemistry A}\ }\textbf {\bibinfo {volume}
  {5}},\ \bibinfo {pages} {13692--13699} (\bibinfo {year} {2017})}\BibitemShut
  {NoStop}%
\bibitem [{\citenamefont {Paul}\ \emph {et~al.}(2019)\citenamefont {Paul},
  \citenamefont {Zhu}, \citenamefont {Chen}, \citenamefont {Qu},\ and\
  \citenamefont {Dai}}]{stability-Pt}%
  \BibitemOpen
  \bibfield  {author} {\bibinfo {author} {\bibfnamefont {Rajib}\ \bibnamefont
  {Paul}}, \bibinfo {author} {\bibfnamefont {Lin}\ \bibnamefont {Zhu}},
  \bibinfo {author} {\bibfnamefont {Hao}\ \bibnamefont {Chen}}, \bibinfo
  {author} {\bibfnamefont {Jia}\ \bibnamefont {Qu}}, \ and\ \bibinfo {author}
  {\bibfnamefont {Liming}\ \bibnamefont {Dai}},\ }\bibfield  {title} {\enquote
  {\bibinfo {title} {Recent advances in carbon-based metal-free
  electrocatalysts},}\ }\href {https://doi.org/10.1002/adma.201806403}
  {\bibfield  {journal} {\bibinfo  {journal} {Advanced Materials}\ }\textbf
  {\bibinfo {volume} {31}},\ \bibinfo {pages} {1806403} (\bibinfo {year}
  {2019})}\BibitemShut {NoStop}%
\bibitem [{\citenamefont {Li}\ and\ \citenamefont
  {Baek}(2019)}]{stability-Pt-2}%
  \BibitemOpen
  \bibfield  {author} {\bibinfo {author} {\bibfnamefont {Changqing}\
  \bibnamefont {Li}}\ and\ \bibinfo {author} {\bibfnamefont {Jong-Beom}\
  \bibnamefont {Baek}},\ }\bibfield  {title} {\enquote {\bibinfo {title}
  {Recent advances in noble metal (pt, ru, and ir)-based electrocatalysts for
  efficient hydrogen evolution reaction},}\ }\href
  {https://doi.org/10.1021/acsomega.9b03550} {\bibfield  {journal} {\bibinfo
  {journal} {ACS omega}\ }\textbf {\bibinfo {volume} {5}},\ \bibinfo {pages}
  {31--40} (\bibinfo {year} {2019})}\BibitemShut {NoStop}%
\end{thebibliography}%

\end{document}